\documentclass{jfm}

\usepackage{graphicx}
\usepackage{subfig}
\usepackage{newtxtext}
\usepackage{newtxmath}
\usepackage{natbib}
\usepackage{float}
\usepackage{makecell}

\usepackage{cite}
\usepackage{mathtools,amsmath,arraycols,array,textcomp}
\usepackage{booktabs,multirow,arydshln}
\usepackage{hyperref,cleveref}
\usepackage{float}

\usepackage{caption}
\usepackage{color} %
\usepackage{arydshln}

\hypersetup{
    colorlinks = true,
    urlcolor   = blue,
    citecolor  = black,
}

\newcommand{\RomanNumeralCaps}[1]
\linenumbers

\title{Effect of thermal fluctuations on spectra and predictability in compressible decaying isotropic turbulence}

\author{Qihan Ma\aff{1},
  Chunxin Yang\aff{1},
  Song Chen\aff{2},
  Kaikai Feng\aff{1},
  Ziqi Cui\aff{1}
  \and Jun Zhang\aff{1}
  \corresp{\email{jun.zhang@buaa.edu.cn}},
  }

\affiliation{\aff{1}School of Aeronautic Science and Engineering, Beihang University, Beijing, 100191, China
             \aff{2}Sino-French Engineer School/School of General Engineering, Beihang University, Beijing 100191, China
            }

\begin{document}
\maketitle

\begin{abstract}
  This study investigates the impact of molecular thermal fluctuations on compressible decaying isotropic turbulence using the unified stochastic particle (USP) method, encompassing both two-dimensional (2D) and three-dimensional (3D) scenarios. The findings reveal that the turbulent spectra of velocity and thermodynamic variables follow the wavenumber scaling law of ${k}^{(d-1)}$ for different spatial dimensions $d$ within the high wavenumber range, indicating the impact of thermal fluctuations on small-scale turbulent statistics. With the application of Helmholtz decomposition, it is found that the thermal fluctuation spectra of solenoidal and compressible velocity components (${\vec{u}}_{s}$ and ${\vec{u}}_{c}$) follow an energy ratio of 1:1 for 2D cases, while the ratio changes to 2:1 for 3D cases. Comparisons between 3D turbulent spectra obtained through USP simulations and direct numerical simulations of the Navier-Stokes equations demonstrate that thermal fluctuations dominate the spectra at length scales comparable to the Kolmogorov length scale. Additionally, the effect of thermal fluctuations on the spectrum of ${\vec{u}}_{c}$ is significantly influenced by variations in the turbulent Mach number. We further study the impact of thermal fluctuations on the predictability of turbulence. With initial differences caused by thermal fluctuations, different flow realizations display significant disparities in velocity and thermodynamic fields at larger scales after a certain period of time, which can be characterized by “inverse error cascades”. Moreover, the results suggest a strong correlation between the predictabilities of thermodynamic fields and the predictability of ${\vec{u}}_{c}$.
\end{abstract}

% \begin{keywords}
% Authors should not enter keywords on the manuscript, as these must be chosen by the author during the online submission process and will then be added during the typesetting process (see \href{https://www.cambridge.org/core/journals/journal-of-fluid-mechanics/information/list-of-keywords}{Keyword PDF} for the full list).  Other classifications will be added at the same time.
% \end{keywords}

% {\bf MSC Codes }  {\it(Optional)} Please enter your MSC Codes here

\section{Introduction}
\label{sec:intro}
According to the classical physical understanding of turbulence, turbulent kinetic energy (TKE) is transferred from the largest scales to successively smaller ones \citep{Pope2000}. The Kolmogorov length scale $\eta$ is the characteristic length scale below which TKE is dominantly dissipated into heat by viscosity. It is defined as $\eta ={{\left( {{{\nu }^{3}}}/{\varepsilon }\; \right)}^{{1}/{4}\;}}$, where $\nu$ is the mean kinematic viscosity and $\varepsilon$ is the mean dissipation rate per unit mass \citep{Pope2000}. The corresponding Kolmogorov time scale ${{\tau }_{\eta }}$ can be defined as ${{\tau }_{\eta }}={{\left( {\nu }/{\varepsilon }\; \right)}^{{1}/{2}\;}}$ \citep{Pope2000}.

As the characteristic scales of turbulence decrease, it becomes crucial to assess the influence of molecular effects on turbulent motions. For a turbulent gas flow characterized by turbulent Reynolds number ${{{Re}}_{T}}$ and turbulent Mach number ${{M}_{t}}$, the ratios of the Kolmogorov scales to the molecular scales can be estimated as \citep{Corrsin1959validityNS,Moser2006validity}
\begin{equation}
  \frac{\eta }{{{\lambda }_{mic}}}={{C}_{1}}\frac{{{{Re}}_{T}}^{{1}/{4}\;}}{{{M}_{t}}},\frac{{{\tau }_{\eta }}}{{{\tau }_{mic}}}={{C}_{2}}\frac{{{{Re}}_{T}}^{{1}/{2}\;}}{{{M}_{t}}^{2}},
  \label{valid_NS}
\end{equation}
where ${\lambda }_{mic}$ and ${{\tau }_{mic}}$ denote the molecular mean free path and the molecular mean collision time, respectively, ${C}_{1}$ and ${C}_{2}$ are two constants of order 1. (\ref{valid_NS}) indicates that for low ${{M}_{t}}$ and high ${{{Re}}_{T}}$, the Kolmogorov scales are considerably larger than the molecular scales. As a result, it is widely believed that the microscopic molecular motions have negligible effects on the macroscopic turbulent motions, and that the Navier-Stokes (NS) equations can accurately describe the turbulent fluctuations at all scales \citep{Khurshid2018dissipationPRF,Buaria2020Dissipation}.

However, several studies have suggested that spontaneous thermal fluctuations \citep{Zhang2009DSMCRB,Ma2021RBS} resulting from molecular motions could have considerable impacts on turbulence. In terms of the statistical properties of turbulence, \citet{Betchov1957ThermalTurb,Betchov1964ThermalTurb} hypothesized that thermal fluctuations could significantly impact the turbulence statistics in the dissipation range. This hypothesis was recently confirmed by \citet{Bell2022thermal}, who numerically solved the incompressible Landau-Lifshitz Navier-Stokes (LLNS) equations of fluctuating hydrodynamics \citep{Landau1959}. These equations incorporate additional stochastic fluxes to model the effect of thermal fluctuations. The study revealed that below length scales comparable to $\eta$, the thermal fluctuations profoundly alter the exponentially decaying turbulent kinetic energy spectrum \citep{Khurshid2018dissipationPRF,Buaria2020Dissipation} predicted by the deterministic NS equations. Additionally, by calculating the probability distribution functions for higher-order derivatives of the velocity, the study reported that the extreme intermittency in the far-dissipation range \citep{Kraichnan1967intermittecy,Chen1993FarDissipation} predicted by the deterministic NS equations is replaced by Gaussian thermal equipartition \citep{Bell2022thermal}. To investigate the effects of thermal fluctuations on turbulence under higher Reynolds number conditions, \citet{Bandak2022thermalPRE} numerically solved the stochastic shell model equations, which can be considered as surrogates of incompressible LLNS equations. They not only revealed the impact of thermal fluctuations on the turbulent energy spectrum in the dissipation range but also investigated the interactions between thermal fluctuations and turbulent intermittency.

Since thermal fluctuations are inherently caused by molecular motions, the molecular simulation methods, such as the molecular dynamics (MD) \citep{Komatsu2014MD_turb,Smith2015MDturbulence} and the direct simulation Monte Carlo (DSMC) \citep{Bird1994DSMC}, can provide a direct way to investigate the role of thermal fluctuations in turbulence. Unlike the simulation methods based on fluctuating hydrodynamics, molecular simulation methods do not assume local thermodynamic equilibrium \citep{Sengers2006fluctuations,McMullen2022NSturbulence}. As a result, they are more suitable for simulating highly compressible turbulence with local nonequilibrium effects.

In recent years, the DSMC method has been extensively employed to simulate compressible turbulent gas flows \citep{Gallis2017turbulencePRL,Gallis2018Couette,Gallis2021turbulence,McMullen2022NSturbulence,McMullen2022porous,McMullen2023thermal,Ma2023thermal}, with several studies focusing on the effect of thermal fluctuations on turbulence statistics. \citet{McMullen2022NSturbulence,McMullen2023thermal} employed the DSMC method to simulate the three-dimensional (3D) Taylor-Green vortex flow, revealing significant influences of thermal fluctuations on both the turbulent energy spectra and velocity structure functions at dissipation length scales. Our recent work \citep{Ma2023thermal} employed DSMC to simulate the two-dimensional (2D) decaying isotropic turbulence, indicating that thermal fluctuations impacted both energy spectra and thermodynamic spectra in the dissipation range. By applying the Helmholtz decomposition \citep{Samtaney2001decay,Wang2012compressibilityJFM} to the 2D velocity field, the effects of thermal fluctuations on the solenoidal and compressible velocity components were studied separately under different ${{M}_{t}}$ conditions \citep{Ma2023thermal}.

In this study, one of our objectives is to explore whether the conclusions we previously drew for 2D cases can be extended to 3D cases. It should be noted that simulating 3D turbulence using DSMC requires a huge computational cost \citep{Gallis2017turbulencePRL,Gallis2021turbulence} due to certain limitations of the method. Specifically, the cell sizes and time steps need to be smaller than ${\lambda }_{mic}$ and ${{\tau }_{mic}}$, respectively \citep{Alexander1998DSMCcellerror,Nicolas2000DSMCcellerror}. To address this challenge, several multiscale particle simulation methods have been proposed \citep{Jenny2010FP,Fei2017FP,Fei2019USP,Fei2021hybridUSP,Fei2023time_relax}. One promising method is the unified stochastic particle (USP) method \Citep{Fei2019USP,Fei2021hybridUSP}. In comparison to DSMC, USP can be implemented with much larger time steps and cell sizes by coupling the effects of molecular movements and collisions. Hence, exploring 3D turbulence through the USP method becomes intriguing, given its inherent inclusion of thermal fluctuations as a particle method and its superior efficiency compared to DSMC.

In addition to influencing turbulent statistics, thermal fluctuations may also play an important role in the predictability of turbulence \citep{Ruelle1979micro,Boffetta2002predict}. Due to the chaotic nature of turbulent flows, even small disturbances in the flow field may lead to the gradual loss of predictability in large-scale turbulent structures over time \citep{Qin2022noise}. The predictability of incompressible turbulence has historically been studied based on the deterministic NS equations \citep{Lorenz1968predict,Leith1972_turb,Olivier1986predict,Boffetta1997predict2D,Boffetta2001predict,Boffetta2017predict,Berera2018chaotic}, focusing on the divergence of velocity field trajectories which initially differ due to artificial perturbations. Given that thermal fluctuations are inherent disturbances in fluids, there is considerable interest in numerically investigating their effects on the predictability of turbulence using particle methods.

In this work, we employ the USP method to simulate compressible decaying isotropic turbulence, aiming to investigate the effects of thermal fluctuations on turbulent spectra and predictability. The rest of the paper is organized as follows. §\,\ref{sec:theory} introduces the basic theories of thermal fluctuations, followed by an overview of the USP method in §\,\ref{sec:method}. In §\,\ref{sec:2D}, the applicability of the USP method is validated by comparing its results with those obtained using the DSMC method for 2D decaying turbulence. Subsequently, in §\,\ref{sec:3D}, the USP method is employed to simulate 3D decaying turbulence. By comparing the results obtained using the USP method with those predicted by the deterministic NS equations \citep{Wang2010hybrid}, the impact of thermal fluctuations on turbulent spectra is studied under different ${{M}_{t}}$ conditions. §\,\ref{sec:predict} discusses the effect of thermal fluctuations on the predictability of turbulence. Conclusions are drawn in §\,\ref{sec:conclusions}.

\section{Spatial correlation of thermal fluctuations}
\label{sec:theory}
In general, the fluctuation of a given macroscopic property $A$ is defined as the difference between its instantaneous local value and its mean value, i.e., $\delta A(\vec{r},t)=A(\vec{r},t)-\left\langle A \right\rangle $ \citep{Pope2000}. In following discussions, we assume that the macroscopic velocity $\vec{u}$ has zero mean, so $\delta\vec{u}=\vec{u}$.

According to the theory of statistical physics, for gases in global thermodynamic equilibrium, the mean square value of the $x$–component velocity fluctuations measured in a volume $V$ is given as \citep{Landau1980equilibrium,Nicolas2003error}
\begin{equation}
  \left\langle {{(u_{x}^{\text{th}})}^{2}} \right\rangle =\frac{{{k}_{B}}\left\langle T \right\rangle }{V\left\langle \rho  \right\rangle },
  \label{eq_velocity}
\end{equation}
where the superscript “th” stands for thermal fluctuations, ${k}_{B}$ is the Boltzmann constant, $\left\langle T \right\rangle$ and $\left\langle \rho  \right\rangle$ are the mean temperature and mass density, respectively. Note that in the equilibrium state, the velocity components are independent and identically distributed, so (\ref{eq_velocity}) also applies to $u_{y}^{\text{th}}$ and $u_{z}^{\text{th}}$ \citep{Landau1980equilibrium}. The total kinetic energy of thermal fluctuations per unit mass is then calculated as
\begin{equation}
  {{K}^{\text{th}}}=\begin{dcases} \begin{array}{*{35}{l}}
    2\text{D}: & \frac{1}{2}\left\langle {{(u_{x}^{\text{th}})}^{2}}+{{(u_{y}^{\text{th}})}^{2}} \right\rangle =\frac{{{k}_{B}}\left\langle T \right\rangle }{V\left\langle \rho  \right\rangle }  \\ [1mm]
    3\text{D}: & \frac{1}{2}\left\langle {{(u_{x}^{\text{th}})}^{2}}+{{(u_{y}^{\text{th}})}^{2}}+{{(u_{z}^{\text{th}})}^{2}} \right\rangle =\frac{3}{2}\frac{{{k}_{B}}\left\langle T \right\rangle }{V\left\langle \rho  \right\rangle }  \\
 \end{array} \end{dcases}.
  \label{eq_kturb}
\end{equation}
For thermal fluctuations of temperature, number density and pressure, their mean square values are given as \citep{Landau1980equilibrium,Nicolas2003error}
\begin{equation}
  T_{ms}^{\text{th}}=\left\langle {{\left( \delta {{T}^{\text{th}}} \right)}^{2}} \right\rangle =\frac{{{k}_{B}}{{\left\langle T \right\rangle }^{2}}}{{{c}_{v}}V\left\langle \rho  \right\rangle },
  \label{eq_Temp}
\end{equation}
\begin{equation}
  n_{ms}^{\text{th}}=\left\langle {{\left( \delta {{n}^{\text{th}}} \right)}^{2}} \right\rangle =\frac{{{\kappa }_{T}}{{k}_{B}}\left\langle T \right\rangle {{\left\langle n \right\rangle }^{\text{2}}}}{V},
  \label{eq_nrho}
\end{equation}
\begin{equation}
  P_{ms}^{\text{th}}=\left\langle {{\left( \delta {{P}^{\text{th}}} \right)}^{2}} \right\rangle =\frac{\gamma {{k}_{B}}\left\langle T \right\rangle }{V{{\kappa }_{T}}},
  \label{eq_Press}
\end{equation}
respectively, where $\left\langle n \right\rangle$ is the mean number density, ${{\kappa}_{T}}={1}/{\left\langle P \right\rangle }\;$ is the isothermal compressibility, $\left\langle P \right\rangle$ is the mean pressure, $\gamma$ denotes the specific heat ratio, and ${{c}_{v}}$ denotes the isochoric specific heat.

For fluctuations satisfying spatial homogeneity, the two-point autocorrelation function $\left\langle \delta A({{{\vec{r}}}_{1}})\delta A({{{\vec{r}}}_{2}}) \right\rangle$ of fluctuations only depends on the relative distance $\vec{l}={{\vec{r}}_{2}}-{{\vec{r}}_{1}}$ \citep{Pope2000}. Providing that $\left| {\vec{l}} \right|$ is much larger than the interatomic distances, the equilibrium thermal fluctuations at different positions are uncorrelated. The two-point autocorrelation functions of $u_{x}^{\text{th}}$, $\delta {{T}^{\text{th}}}$, $\delta {{n}^{\text{th}}}$ and $\delta {{P}^{\text{th}}}$ are given as \citep{Landau1980condensed}
\begin{equation}
  \mathcal{R}_{{{u}_{x}}}^{\text{th}}(\vec{l})=\left\langle u_{x}^{\text{th}}({{{\vec{r}}}_{1}})u_{x}^{\text{th}}({{{\vec{r}}}_{2}}) \right\rangle =\frac{{{k}_{B}}\left\langle T \right\rangle }{\left\langle \rho  \right\rangle }\delta (\vec{l}),
  \label{cor_velocity}
\end{equation}
\begin{equation}
  \mathcal{R}_{T}^{\text{th}}(\vec{l})=\left\langle \delta {{T}^{\text{th}}}({{{\vec{r}}}_{1}})\delta {{T}^{\text{th}}}({{{\vec{r}}}_{2}}) \right\rangle =\frac{{{k}_{B}}{{\left\langle T \right\rangle }^{2}}}{{{c}_{v}}\left\langle \rho  \right\rangle }\delta (\vec{l}),
  \label{cor_Temp}
\end{equation}
\begin{equation}
  \mathcal{R}_{n}^{\text{th}}(\vec{l})=\left\langle \delta {{n}^{\text{th}}}({{{\vec{r}}}_{1}})\delta {{n}^{\text{th}}}({{{\vec{r}}}_{2}}) \right\rangle ={{\kappa }_{T}}{{k}_{B}}\left\langle T \right\rangle {{\left\langle n \right\rangle }^{\text{2}}}\delta (\vec{l}),
  \label{cor_nrho}
\end{equation}
\begin{equation}
  \mathcal{R}_{P}^{\text{th}}(\vec{l})=\left\langle \delta {{P}^{\text{th}}}({{{\vec{r}}}_{1}})\delta {{P}^{\text{th}}}({{{\vec{r}}}_{2}}) \right\rangle =\frac{\gamma {{k}_{B}}\left\langle T \right\rangle }{{{\kappa }_{T}}}\delta (\vec{l}),
  \label{cor_Press}
\end{equation}
respectively, where $\delta(\vec{l})$ denotes the Dirac delta function. In (\ref{cor_velocity})–(\ref{cor_Press}), we have taken the limit as the volume $V$ approaches zero.

The energy spectrum $E(k)$ can be expressed as the Fourier transform of the two-point velocity autocorrelation function \citep{Pope2000}:
\begin{equation}
  E(k)=E(\left| {\vec{k}} \right|)=\begin{dcases} \begin{array}{*{35}{l}}
    2\text{D}: & \frac{1}{2}\left( \mathcal{F}\left\{ {{\mathcal{R}}_{{{u}_{x}}}} \right\}+\mathcal{F}\left\{ {{\mathcal{R}}_{{{u}_{y}}}} \right\} \right)\times 2\pi k  \\
    3\text{D}: & \frac{1}{2}\left( \mathcal{F}\left\{ {{\mathcal{R}}_{{{u}_{x}}}} \right\}+\mathcal{F}\left\{ {{\mathcal{R}}_{{{u}_{y}}}} \right\}+\mathcal{F}\left\{ {{\mathcal{R}}_{{{u}_{z}}}} \right\} \right)\times 4\pi {{k}^{2}}  \\
    \end{array} \end{dcases},
\label{energy_spec}
\end{equation}
where $\mathcal{F}\left\{ A \right\}=\int\limits_{-\infty }^{+\infty }{A(\vec{r}){{e}^{-i\vec{k}\cdot \vec{r}}}}d\vec{r}$ denotes the spatial Fourier transform of $A$ with respect to the wave vector $\vec{k}$. The terms $2\pi k$ and $4\pi{{k}^{2}}$ appear in (\ref{energy_spec}) due to the integration of the spectrum over the wavenumber circle or sphere surface in 2D or 3D cases \citep{Verma2020equilibrium}. By substituting (\ref{cor_velocity}) into (\ref{energy_spec}), one can yield the energy spectrum of thermal fluctuations as
\begin{equation}
{{E}^{\text{th}}}(k)=\begin{dcases} \begin{array}{*{35}{l}}
  2\text{D}: & \frac{{{k}_{B}}\left\langle T \right\rangle }{\left\langle \rho  \right\rangle }\times 2\pi k  \\ [1mm]
  3\text{D}: & \frac{3}{2}\frac{{{k}_{B}}\left\langle T \right\rangle }{\left\langle \rho  \right\rangle }\times 4\pi {{k}^{2}}  \\
\end{array} \end{dcases}.
\label{energy_spec_eq}
\end{equation}
Therefore, it can be concluded that for gases in equilibrium, the 2D energy spectrum grows linearly with the wavenumber $k$ \citep{Ma2023thermal}, while the 3D energy spectrum grows quadratically with $k$ \citep{Bell2022thermal,McMullen2022NSturbulence,Bandak2022thermalPRE}.

Similarly, the spectra of fluctuating thermodynamic variables can be expressed as
\begin{equation}
  {{E}_{g}}(k)=\begin{dcases} \begin{array}{*{35}{l}}
    2\text{D}: & \mathcal{F}\left\{ {{\mathcal{R}}_{g}} \right\}\times 2\pi k  \\
    3\text{D}: & \mathcal{F}\left\{ {{\mathcal{R}}_{g}} \right\}\times 4\pi {{k}^{2}}  \\
\end{array} \end{dcases},
\label{thermal_spec}
\end{equation}
where $g$ represents the temperature $T$, number density $n$ or pressure $P$. Substituting (\ref{cor_Temp})–(\ref{cor_Press}) into (\ref{thermal_spec}) leads to the same conclusion that the equilibrium spectra of thermodynamic variables grow linearly with $k$ for 2D cases, while they grow quadratically with $k$ for 3D cases.

\begin{figure}
  \centering
  \includegraphics[scale=0.6]{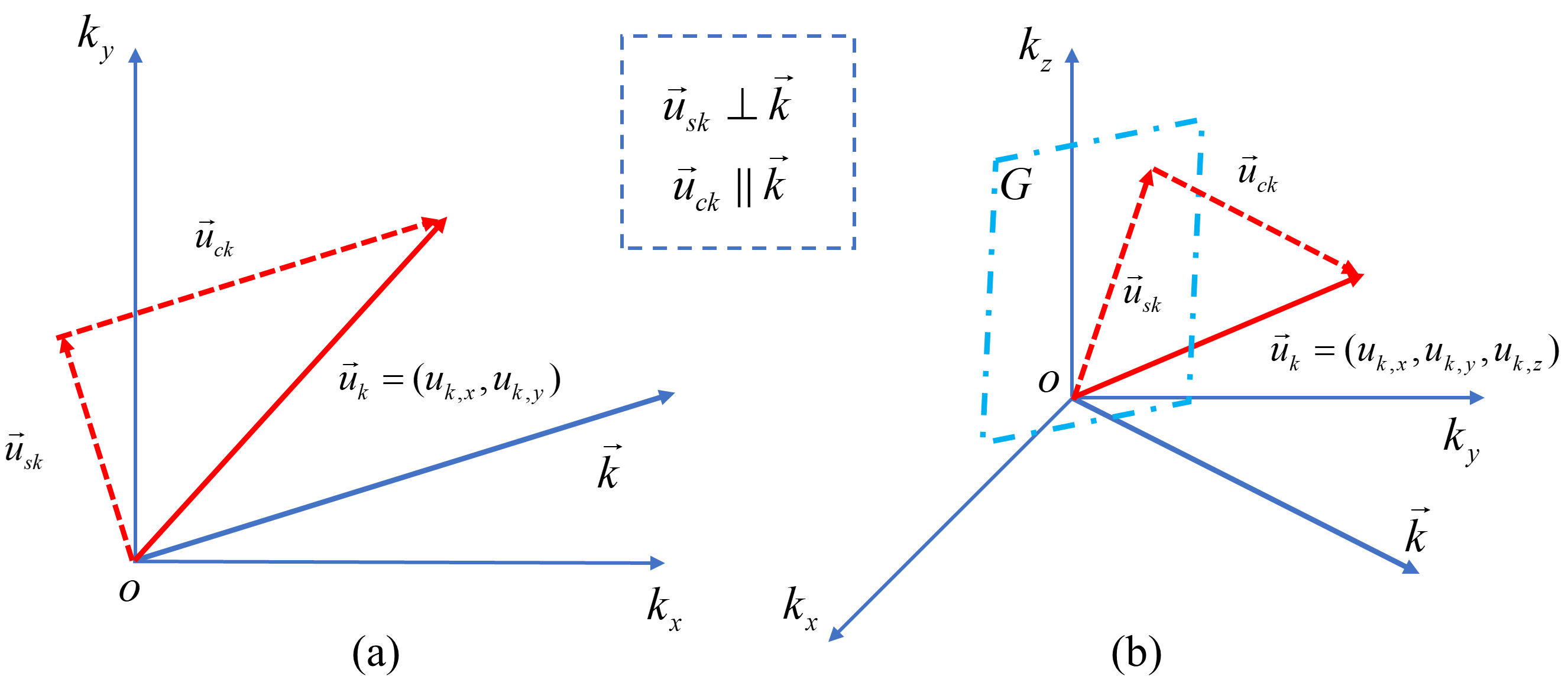}
  \caption{Sketch (in wavenumber space) showing the decomposition of the fluctuating velocity field ${{\vec{u}}_{k}}=\left( {{u}_{k,i}} \right)$ into the solenoidal component ${{\vec{u}}_{sk}}$ and the compressible component ${{\vec{u}}_{ck}}$, for 2D (a) and 3D (b) cases. In panel (b), ${{\vec{u}}_{sk}}$ lies in the plane $G$, which is perpendicular to $\vec{k}$.}
\label{fig1}
\end{figure}

For compressible fluids, the Helmholtz decomposition \citep{Samtaney2001decay,Wang2012compressibilityJFM} is always applied to the fluctuating velocity field as $\vec{u}={{\vec{u}}_{s}}+{{\vec{u}}_{c}}$, where the solenoidal component ${{\vec{u}}_{s}}$ and the compressible component ${{\vec{u}}_{c}}$ satisfy conditions $\nabla \cdot {{\vec{u}}_{s}}=0$ and $\nabla \times {{\vec{u}}_{c}}=0$, respectively. In wavenumber space, the Helmholtz decomposition can be applied as \citep{Pope2000}
\begin{equation}
  {{\vec{u}}_{ck}}={\vec{k}(\vec{k}\cdot {{{\vec{u}}}_{k}})}/{{{k}^{2}}},
\label{helm_c}
\end{equation}
\begin{equation}
  {{\vec{u}}_{sk}}={{\vec{u}}_{k}}-{{\vec{u}}_{ck}},
\label{helm_s}
\end{equation}
where ${{\vec{u}}_{k}}$, ${{\vec{u}}_{ck}}$ and ${{\vec{u}}_{sk}}$ denote the spatial Fourier transform of $\vec{u}$, ${{\vec{u}}_{c}}$ and ${{\vec{u}}_{s}}$, respectively. (\ref{helm_c})–(\ref{helm_s}) indicate that ${{\vec{u}}_{sk}}$ is perpendicular to $\vec{k}$, while ${{\vec{u}}_{ck}}$ is parallel to $\vec{k}$ (see figure \ref{fig1}).

To calculate the energy spectra of $\vec{u}_{c}^{\text{th}}$ and $\vec{u}_{s}^{\text{th}}$, note that in wavenumber space, each independent velocity component $u_{k,i}^{\text{th}}$ shares the same amount of energy, given as
\begin{equation}
  {{\left| u_{k,i}^{\text{th}} \right|}^{2}}=\mathcal{F}\left\{ \mathcal{R}_{{{u}_{i}}}^{\text{th}} \right\}={{{k}_{B}}\left\langle T \right\rangle }/{\left\langle \rho \right\rangle }\;.
  \label{helm_energy}
\end{equation}
Therefore, it follows that ${{\left| {{{\vec{u}}}^{\text{th}}}_{sk} \right|}^{2}}={{\left| {{{\vec{u}}}^{\text{th}}}_{ck} \right|}^{2}}$ for 2D cases, while ${{\left| {{{\vec{u}}}^{\text{th}}}_{sk} \right|}^{2}}=2{{\left| {{{\vec{u}}}^{\text{th}}}_{ck} \right|}^{2}}$ for 3D cases. The energy spectra of $\vec{u}_{c}^{\text{th}}$ and $\vec{u}_{s}^{\text{th}}$ can then be calculated from ${{E}^{\text{th}}}(k)$ as
\begin{equation}
  \begin{dcases} \begin{array}{*{35}{l}}
  2\text{D}: & E_{c}^{\text{th}}(k)=E_{s}^{\text{th}}(k)=\frac{1}{2}{{E}^{\text{th}}}(k)  \\
  3\text{D}: & E_{c}^{\text{th}}(k)=\frac{1}{2}E_{s}^{\text{th}}(k)=\frac{1}{3}{{E}^{\text{th}}}(k)  \\
\end{array} \end{dcases}.
\label{helm_spec}
\end{equation}

\section{Simulation method}
\label{sec:method}
In this work, the unified stochastic particle (USP) method is employed to simulate the compressible decaying isotropic turbulence. Here we provide a brief description of the theoretical background and the basic algorithm of USP, and we refer readers to the original papers \citep{Fei2019USP,Fei2021hybridUSP,Fei2021USP_AIA} for details.

\subsection{Governing equations}
\label{subsec:equation}
According to the gas kinetic theory, the state of a gas can be described by the velocity distribution function (VDF) $f(\vec{c};\vec{r},t)$, which is defined as the number density of molecules with velocity $\vec{c}$ at position $\vec{r}$ and time $t$. The evolution of VDF can be described by the Boltzmann equation \citep{Bird1994DSMC}:
\begin{equation}
  \frac{\partial f}{\partial t}+\vec{c}\cdot \nabla f=Q\left( f,{{f}^{*}} \right),
  \label{Boltzmann}
\end{equation}
where the term $\vec{c}\cdot\nabla f$ describes the change of VDF due to the convection of molecules, and $Q\left( f,{{f}^{*}} \right)$ is an integral that describes the intermolecular collisions. Due to the challenges associated with directly solving the Boltzmann equation, most numerical works are based on its model equations like the Bhatnagar-Gross-Krook (BGK) model \citep{BGK1954} or the Shakhov-BGK (S-BGK) model \citep{Shakhov1968SBGK}. These models simplify the Boltzmann collision integral with a linear relaxation term, i.e.,
\begin{equation}
  \frac{\partial f}{\partial t}+\vec{c}\cdot \nabla f=\frac{{{f}_{t}}-f}{{{\tau }_{r}}},
  \label{BGK}
\end{equation}
where the right-hand side of (\ref{BGK}) describes the relaxation of VDF towards a target distribution function ${{f}_{t}}$ with the relaxation time ${{\tau }_{r}}$ comparable to the molecular mean collision time ${{\tau }_{mic}}$. In BGK and S-BGK models, the target distribution functions are given by the local macroscopic quantities as \citep{Yao2023ExtendSBGK}
\begin{equation}
  {{f}_{t}}^{\text{BGK}}={{f}_{M}}=n{{\left( \frac{1}{2\pi RT} \right)}^{\frac{3}{2}}}\exp \left( -\frac{{{C}^{2}}}{2RT} \right),
  \label{f_BGK}
\end{equation}
\begin{equation}
  {{f}_{t}}^{\text{S-BGK}}={{f}_{M}}\left[ 1+(1-Pr)\frac{{{C}_{i}}{{q}_{i}}}{5PRT}\left( \frac{{{C}^{2}}}{2RT}-\frac{5}{2} \right) \right],
  \label{f_SBGK}
\end{equation}
where $\vec{C}=\vec{c}-\vec{u}$ is the molecular thermal velocity, $R$ is the specific gas constant, $Pr$ is the Prandtl number, and ${{q}_{i}}$ is the heat flux. Compared with the original BGK model with a fixed $Pr$ of 1, the S-BGK model can be applied to gas flows with arbitrary $Pr$ \citep{Yao2023ExtendSBGK}.

\subsection{Unified stochastic particle method}
\label{subsec:USP}
So far, the direct simulation Monte Carlo (DSMC) method \citep{Bird1994DSMC} is still the most commonly used molecular method for simulating rarefied gas flows, and it has recently been employed to investigate the effect of thermal fluctuations on turbulence \citep{McMullen2022NSturbulence,McMullen2023thermal,Ma2023thermal}. A typical DSMC simulation tracks an appropriate number of “particles” (simulated molecules) in the computational domain. Each particle statistically represents a fixed number $F$ of identical real molecules, and $F$ is the so-called simulation ratio \citep{Gallis2017turbulencePRL,McMullen2022NSturbulence}. The domain is divided into computational cells where local macroscopic quantities are obtained by sampling particle information.

The key point of the DSMC method is that the effects of molecular movements and collisions are assumed uncoupled within a computational time step $\Delta t$. Specifically, the simulated particles move ballistically first, then the particles within the same cell are randomly chosen as collision pairs to assign new velocities according to the phenomenological collision models \citep{Bird1994DSMC}. DSMC can be regarded as an operator splitting scheme to solve the Boltzmann equation \citep{Wagner1992Convergence,Kai2023USP_CPC}, i.e.,
\begin{equation}
  \begin{array}{*{35}{l}}
     \frac{\partial f}{\partial t}=-\vec{c}\cdot \nabla f, \\ [1mm]
     \frac{\partial f}{\partial t}=Q\left( f,{{f}^{*}} \right). \\ 
  \end{array}
\end{equation}

The same procedure can also be applied to (\ref{BGK}), resulting in the governing equations of the stochastic particle (SP) method based on the BGK model \citep{Gallis2000SP,Macrossan2001SP,Pfeiffer2018SPBGK}, given by
\begin{equation}
  \begin{array}{*{35}{l}}
    \frac{\partial f}{\partial t}=-\vec{c}\cdot \nabla f, \\ [1mm]
    \frac{\partial f}{\partial t}=\frac{{{f}_{t}}-f}{{{\tau }_{r}}}. \\ 
  \end{array}
\end{equation}
In the SP method, the process of molecular movements is the same as that in the DSMC method, while the process of intermolecular collisions in DSMC is replaced by a “redistribution phase” where a fraction $(1-\exp \left( -{\Delta t}/{{{\tau }_{r}}}\; \right))$ of particles in each cell are randomly selected to assign new velocities according to ${{f}_{t}}$. The velocities of the remaining fraction of particles are unchanged.

Theoretically, it has been proved that DSMC and SP will produce unphysical momentum and energy transport if the time step $\Delta t$ and cell size $\Delta {{L}_{cell}}$ exceed ${{\tau }_{mic}}$ and ${{\lambda }_{mic}}$, respectively \citep{Alexander1998DSMCcellerror,Nicolas2000DSMCcellerror,Fei2019USP}. To address this issue, the USP method supplements the effect of intermolecular collisions in the convection step. The corresponding governing equations based on the S-BGK model can be written as
\begin{equation}
  \begin{array}{*{35}{l}}
     \frac{\partial f}{\partial t}=-\vec{c}\cdot \nabla f+{Q}^{*} , \\ [1mm]
     \frac{\partial f}{\partial t}=\frac{{{f}_{t}}^{S-BGK}-f}{{{\tau }_{r}}}-{Q}^{*} , \\ 
  \end{array}
  \label{USP_origin}
\end{equation}
where ${Q}^{*}$ is a modified collision term closed by the Grad’s 13 moment distribution function \citep{Fei2019USP}. To make the USP method easier to be implemented, (\ref{USP_origin}) can be further rewritten as \citep{Fei2021hybridUSP}
\begin{equation}
  \begin{array}{*{35}{l}}
    \frac{\partial f}{\partial t}=-\vec{c}\cdot \nabla f, \\ [1mm]
    \frac{\partial f}{\partial t}=\frac{{{f}_{U}}-f}{{{\tau }_{r}}}, \\ 
  \end{array}
  \label{USP_final}
\end{equation}
where ${{f}_{U}}$ is a new target distribution function given as
\begin{equation}
  {{f}_{U}}={{f}_{M}}\left[ 1+{{\varPsi }_{1}}\frac{{{\sigma }_{ij}}{{C}_{<i}}{{C}_{j>}}}{2PRT}+{{\varPsi }_{2}}\frac{2{{C}_{i}}{{q}_{i}}}{5PRT}\left( \frac{{{C}^{2}}}{2RT}-\frac{5}{2} \right) \right],
\end{equation}
where ${{\varPsi}_{1}}$ and ${{\varPsi}_{2}}$ are related to $\Delta t$ as ${{\varPsi}_{1}}=1-{\Delta t}/{2{{\tau }_{r}}\coth \left( {\Delta t}/{2{{\tau }_{r}}}\right)}$ and ${{\varPsi}_{2}}=1-{Pr \Delta t}/{2{{\tau }_{r}}\coth \left( {\Delta t}/{2{{\tau }_{r}}}\right)}$, respectively. Based on (\ref{USP_final}), it follows that the implementation of USP is quite similar to that of SP. Theoretically, it has been demonstrated that the USP method has second-order temporal accuracy when $\Delta t\gg {{\tau }_{mic}}$ \citep{Fei2021hybridUSP}. Furthermore, the second-order spatial accuracy can be achieved by a spatial interpolation procedure for macroscopic variables \citep{Fei2021USP_AIA}.

In this work, we simulate turbulent flows of the argon gas with $Pr ={2}/{3}$ and $\gamma ={5}/{3}$. The dynamic viscosity $\mu$ is assumed to depend on the temperature with a power-law exponent $\omega$, i.e.,
\begin{equation}
  \mu ={{\mu}_{ref}}{{\left( \frac{T}{{{T}_{ref}}} \right)}^{\omega }},
\end{equation}
where ${{\mu }_{ref}}$ is the reference viscosity at the reference temperature ${{T}_{ref}}$. Specifically for argon gas, $\omega$, ${{\mu }_{ref}}$ and ${{T}_{ref}}$ are set to 0.81, $2.117\times {{10}^{-5}}\;\text{Pa}\cdot \text{s}$ and $273.15\;\text{K}$, respectively \citep{Bird1994DSMC}. The USP simulations are performed using the open-source code SPARTACUS \citep{Kai2023USP_CPC}, which has been recently developed by the authors within the framework of a widely used DSMC solver SPARTA \citep{Plimpton2019spartaPOF}. The performance of SPARTACUS has been evaluated over a series of test cases covering 1D to 3D flows with a wide range of Knudsen numbers and Mach numbers \citep{Kai2023USP_CPC}.

\section{Two-dimensional turbulence}
\label{sec:2D}
In a recent study \citep{Ma2023thermal}, we employed the DSMC method to investigate the effect of thermal fluctuations on the spectra of 2D decaying isotropic turbulence. In this section, we use the DSMC results as benchmarks to validate the applicability of the USP method. The simulations begin with argon gas flows at ${{T}_{0}}=300\;\text{K}$ and ${{P}_{0}}=1\;\text{bar}$, with the number density calculated as ${{n}_{0}}={{{P}_{0}}}/{\left( {{k}_{B}}{{T}_{0}} \right)}$. Based on these initial conditions, the molecular mean collision time ${{\tau }_{mic0}}$ and the molecular mean free path ${{\lambda }_{mic0}}$ are estimated using the variable hard sphere (VHS) model parameters specific to argon \citep{Bird1994DSMC}. The side lengths of the simulation domain are set to $\left( {{L}_{x}},{{L}_{y}},{{L}_{z}} \right)=\left( 4000{{\lambda }_{mic0}},4000{{\lambda }_{mic0}},40{{\lambda }_{mic0}} \right)$, and the domain is divided into uniform computational cells along the $x$ and $y$ directions for 2D simulations.

The initial turbulent velocity field is generated as follows. First, a divergence-free velocity field $\vec{u}_{0}^{\text{NS}}$ with a prescribed energy spectrum is randomly generated using the transfer procedures provided by \citet{Ishiko2009LESturb2D}. The initial energy spectrum is specified as
\begin{equation}
  {{E}^{\text{NS}}}(k,t=0)=\frac{{{a}_{s}}}{2}\frac{U_{0}^{2}}{{{k}_{p}}}{{\left( \frac{k}{{{k}_{p}}} \right)}^{2s+1}}\text{exp}\left[ -(s+\frac{1}{2}){{\left( \frac{k}{{{k}_{p}}} \right)}^{2}} \right],{{a}_{s}}=\frac{{{(2s+1)}^{s+1}}}{{{2}^{s}}s!},
\label{initial_2D_spec}
\end{equation}
where ${{U}_{0}}={{\left\langle {{\left( \vec{u}_{0}^{\text{NS}} \right)}^{2}} \right\rangle }^{0.5}}$ is the root mean square value of $\vec{u}_{0}^{\text{NS}}$, $s$ is a shape parameter of the spectrum, and ${{k}_{p}}$ is the wavenumber at which the spectrum has peak value. In this work, we take $s=3$ and ${{k}_{p}}=9{{k}_{\text{min}}}$, where ${{k}_{\text{min}}}={2\pi }/{L}\;$ is the minimum wavenumber, and $L={{L}_{x}}={{L}_{y}}$. Based on (\ref{initial_2D_spec}), the initial enstrophy is calculated as ${{\varOmega}_{0}}=\int_{0}^{\infty }{{{k}^{2}}{{E}^{\text{NS}}}(k)dk}$ \citep{Herring1974DecayTurb2D}. The enstrophy dissipation rate and the corresponding dissipation length scale are calculated as ${{\varepsilon }_{\varOmega 0}}=2{{\nu }_{0}}\int_{0}^{\infty }{{{k}^{4}}{{E}^{\text{NS}}}(k)dk}$ and ${{\eta }_{\varOmega 0}}={{\left( {{{\nu }_{0}}^{3}}/{{{\varepsilon }_{\varOmega 0}}}\; \right)}^{{1}/{6}\;}}$, respectively \citep{Herring1974DecayTurb2D}. The initial turbulent Mach number and the Taylor Reynolds number are given by
\begin{equation}
  {{M}_{t0}}=\frac{{{U}_{0}}}{\left\langle \sqrt{\gamma RT} \right\rangle },R{{e}_{\lambda 0}}={{{\varOmega }_{0}}^{1.5}}/{{{\varepsilon }_{\varOmega 0}}}\;,
\label{Mach_Re_2D}
\end{equation}
respectively.

The macroscopic velocity ${{\vec{u}}_{0}}^{\text{NS}}$ generated for each computational cell can be considered as the initial solution of deterministic NS equations without thermal fluctuations. The velocities ${{\vec{c}}_{0}}$ of USP particles in each cell are then generated based on the relation ${{\vec{c}}_{0}}={{\vec{u}}_{0}}^{\text{NS}}+{{\vec{C}}_{0}}$, where the particle thermal velocities ${{\vec{C}}_{0}}$ are randomly sampled from the Maxwell distribution function at $\left( {{T}_{0}},{{n}_{0}} \right)$. This procedure enables the initial velocity field in the USP simulation to be expressed as $\vec{u}_{0}^{\text{USP}}=\vec{u}_{0}^{\text{NS}}+\vec{u}_{0}^{\text{th}}$ \citep{McMullen2023thermal}, where $\vec{u}_{0}^{\text{th}}$ represents the thermal velocity fluctuation measured at each cell.

In this section, all simulation cases commence with the same turbulent velocity field with ${{M}_{t0}}=1$ and $R{{e}_{\lambda 0}}=23.3$. The other simulated parameters are shown in table \ref{2Dparameter}. The DSMC simulation is conducted using SPARTA with $\Delta t=0.2{{\tau }_{mic0}}$ and $\Delta {{L}_{cell}}=0.49{{\lambda }_{mic0}}$. In contrast, the USP simulations are conducted with larger $\Delta t$ and $\Delta {{L}_{cell}}$. The average number of simulated particles within each cell ($\left\langle {{N}_{p}} \right\rangle$) increases with $\Delta {{L}_{cell}}$ to maintain the total number of particles unchanged, resulting in the same simulation ratio of $F=1549$. Based on $\Delta {{L}_{cell}}$, we further calculate the resolution parameter ${{k}_{\max }}{{\eta }_{{{\varOmega}_{0}}}}$, where ${{k}_{\max }}={\pi }/{\Delta {{L}_{cell}}}={\pi {{N}_{c}}}/{L}$ denotes the largest wavenumber corresponding to the half of ${{N}_{c}}$ \citep{Wang2017citPRF}. Each simulation case is run on 1024 CPU cores with the total computation time shown in table \ref{2Dparameter}, corresponding to the same physical time of $t=10.7{{t}_{0}}$, where ${{t}_{0}}={L}/{\left( \text{2}\pi {{U}_{0}} \right)}$. Compared to the DSMC method, the USP method shows a significant improvement in computation efficiency due to the increases in $\Delta t$ and $\Delta {{L}_{cell}}$.

\begin{table}
  \centering
    \begin{tabular}{*{7}{c}}
    %\hline
    Case & Resolution (${{N}_{c}}^{2}$) & $\left\langle {{N}_{p}} \right\rangle$ & ${\Delta t}/{{{\tau }_{mic0}}}\;$ &  ${\Delta {{L}_{cell}}}/{{{\lambda }_{mic0}}}\;$ & ${{k}_{\max }}{{\eta }_{\varOmega  0}}$ & Total computation time (hours) \\
    \midrule
    DSMC   &${{8192}^{2}}$	&25	  &0.2	&0.49	 &94.3	&31.00  	   \\
    USP    &${{4096}^{2}}$	&100	&0.5	&0.98	 &47.1	&10.50 	     \\
    USP    &${{2048}^{2}}$	&400	&1.0	&1.95	 &23.6	&4.97 	     \\
    USP    &${{1024}^{2}}$	&1600	&2.0	&3.91	 &11.8	&2.79  	     \\\hline
    \end{tabular}
    \caption{Simulated parameters for 2D decaying isotropic turbulence. All the simulations are performed with the initial conditions of ${{T}_{0}}=300\;\text{K}$, ${{P}_{0}}=1\;\text{bar}$, ${{M}_{t0}}=1$ and $R{{e}_{\lambda 0}}=23.3$.}
  \label{2Dparameter}
\end{table}

\begin{figure}
  \centering
  \subfloat{
      \includegraphics[scale=0.42]{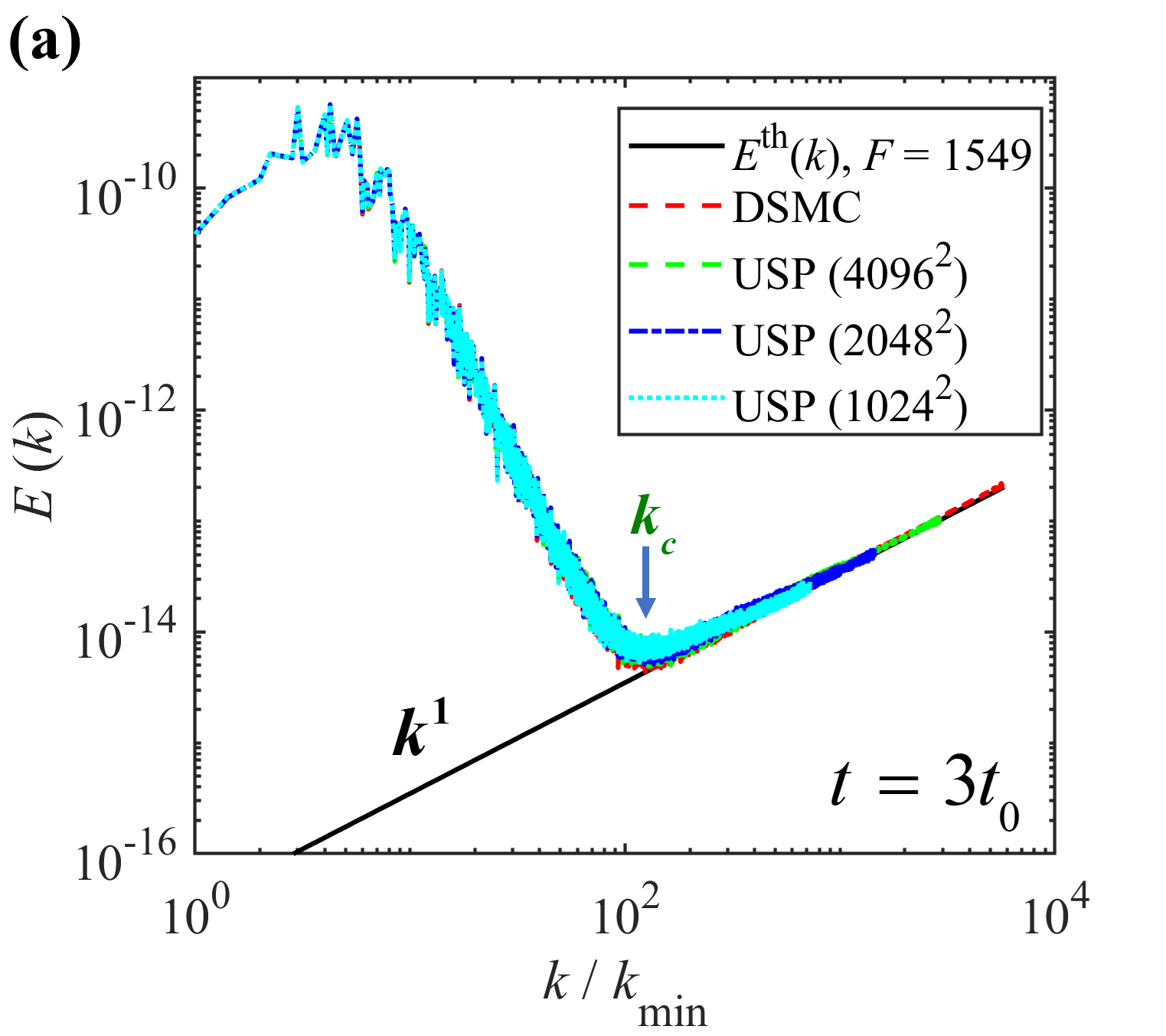}}
  \subfloat{
      \includegraphics[scale=0.42]{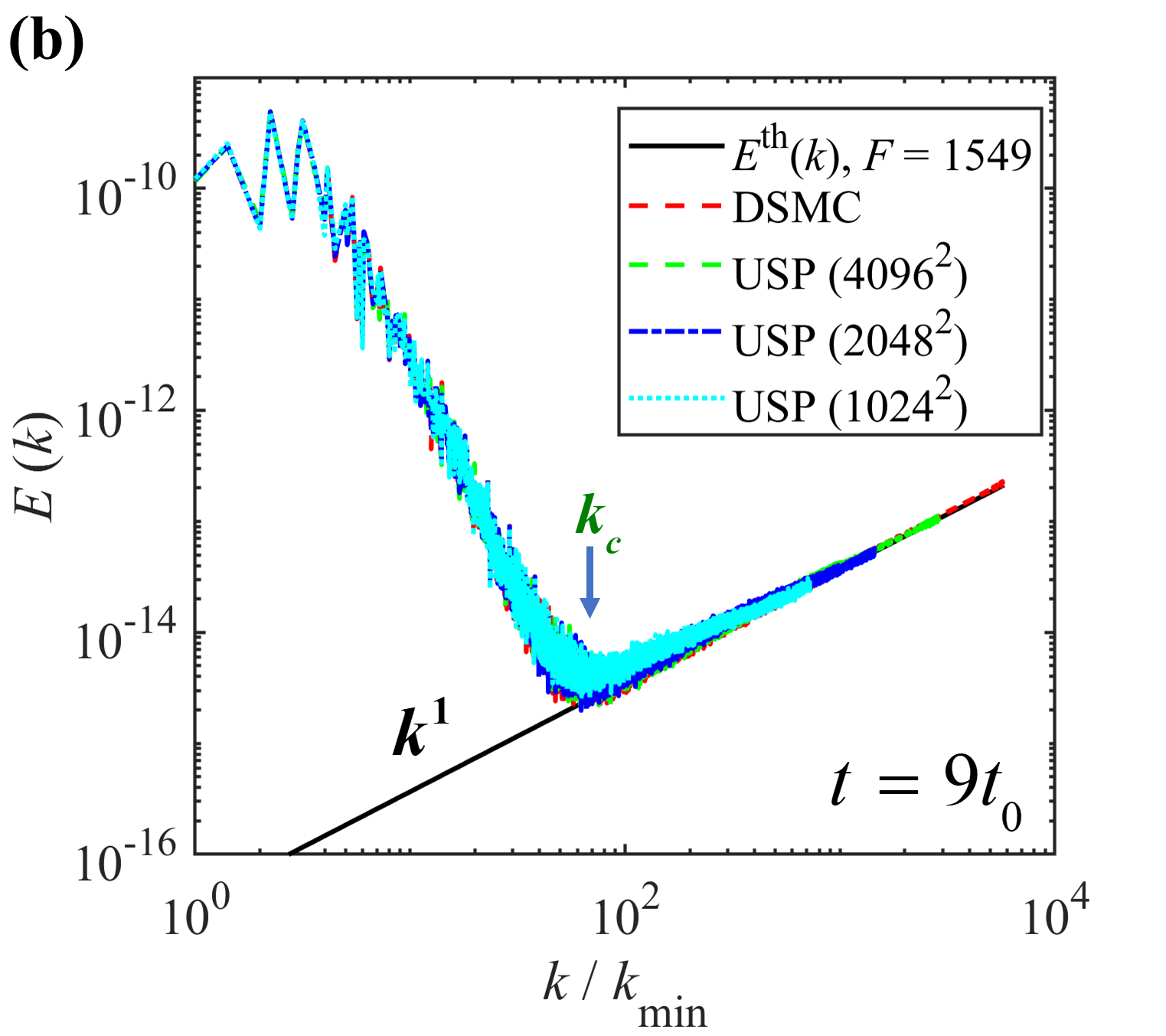}}\\
  \vspace{-1em}
  \subfloat{
      \includegraphics[scale=0.42]{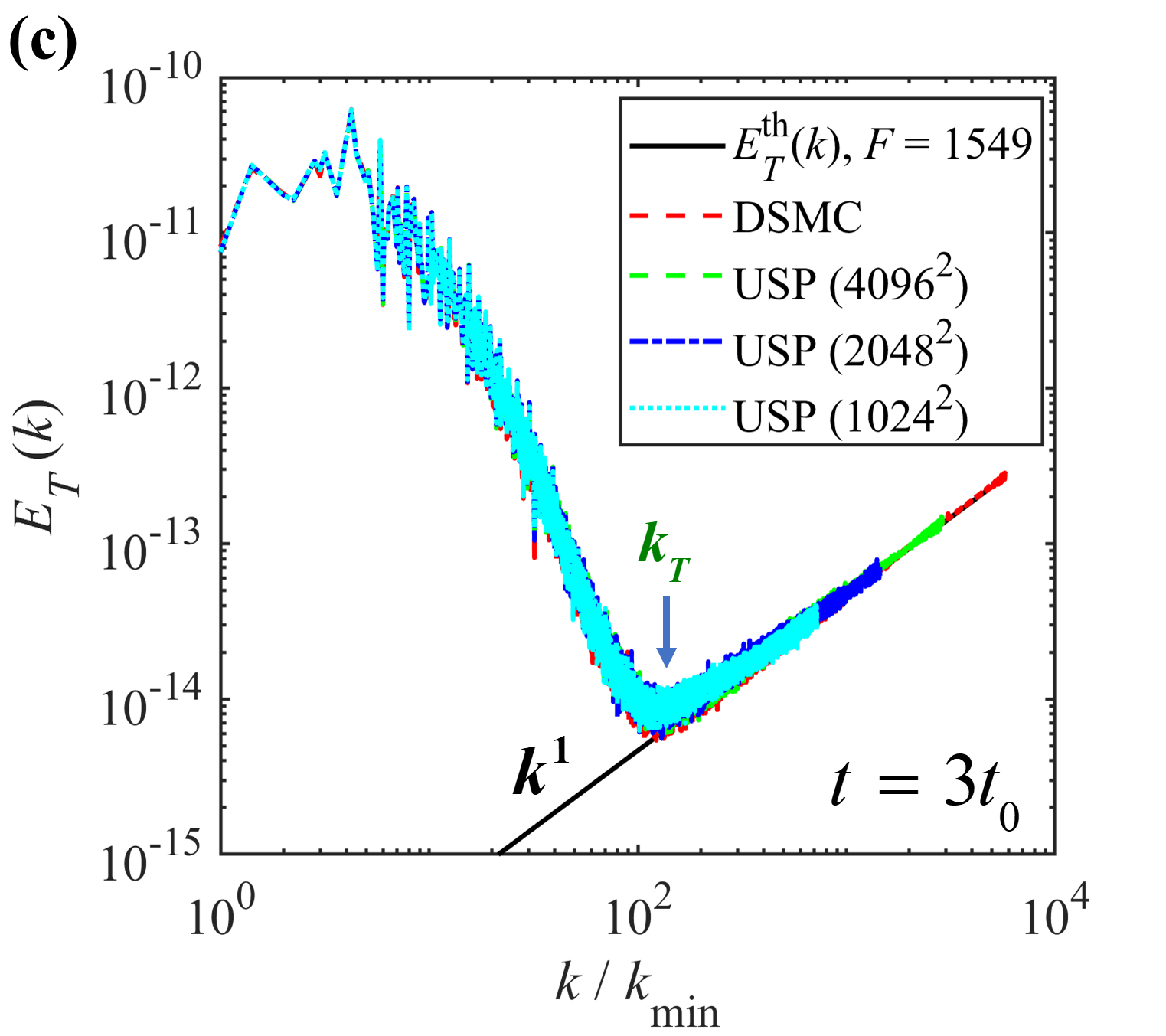}}
  \subfloat{
      \includegraphics[scale=0.42]{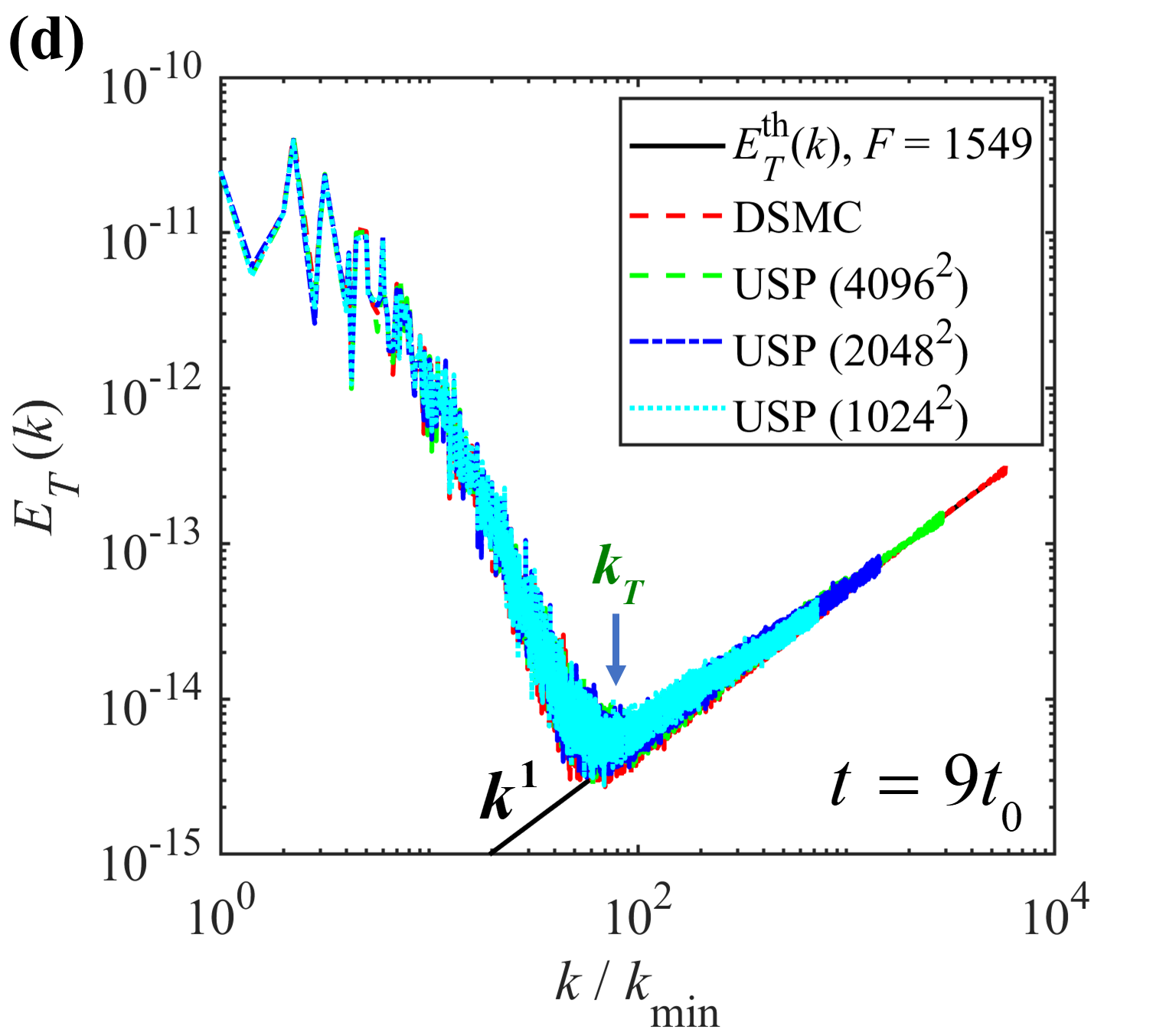}}\\
  \vspace{-1em}
  \subfloat{
      \includegraphics[scale=0.42]{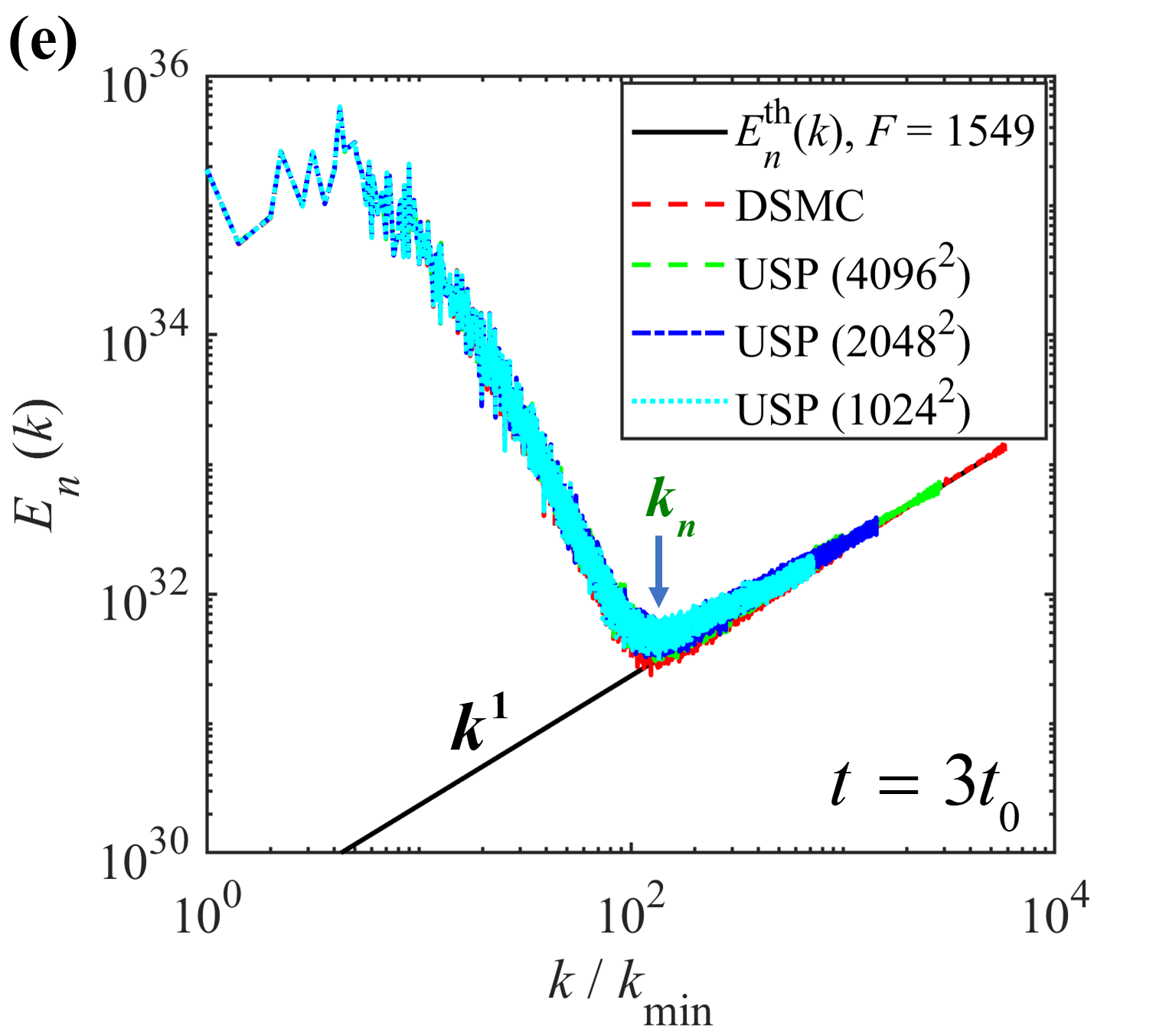}}
  \subfloat{
        \includegraphics[scale=0.42]{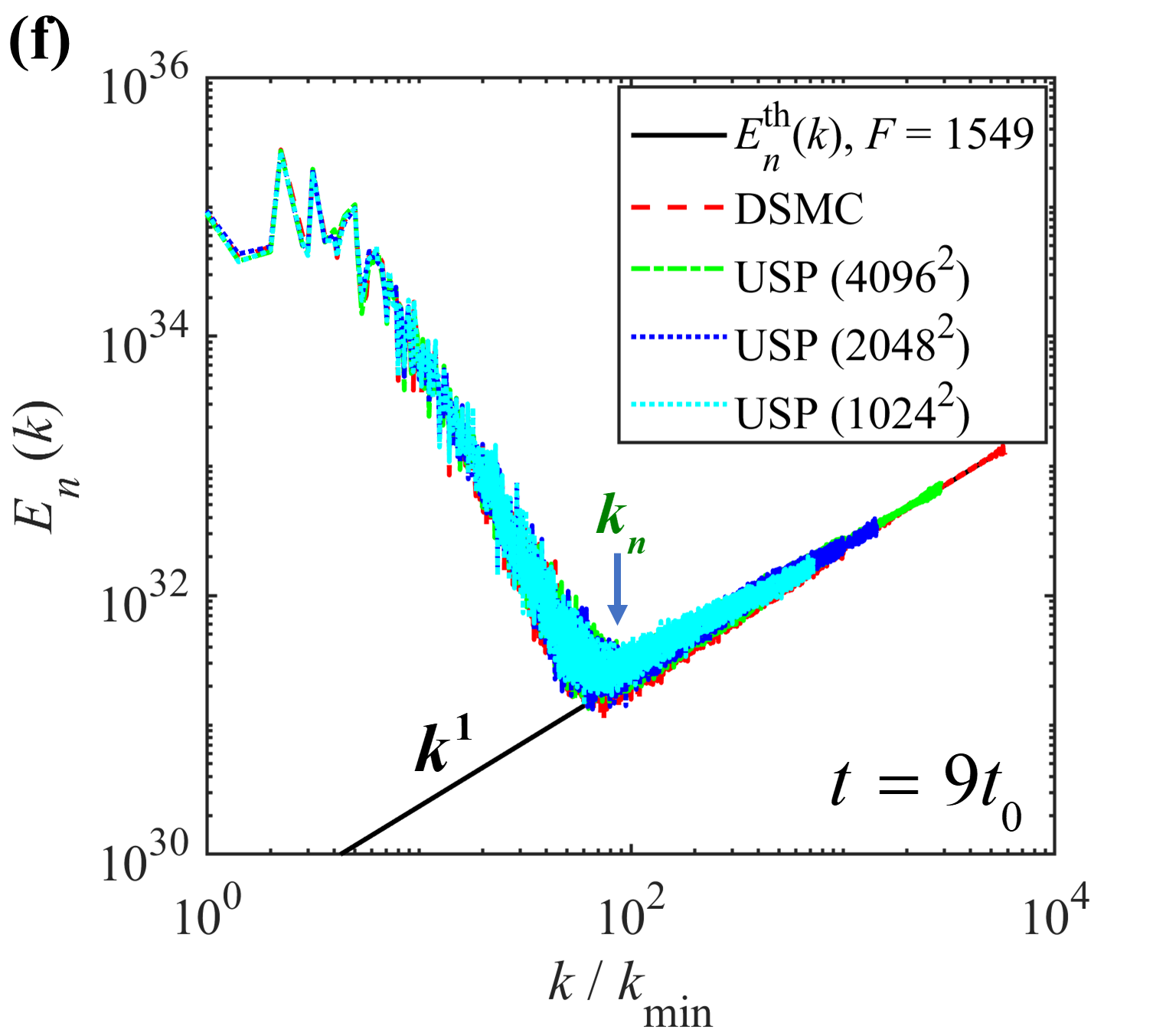}}\\
  \vspace{-1em}
  \subfloat{
        \includegraphics[scale=0.42]{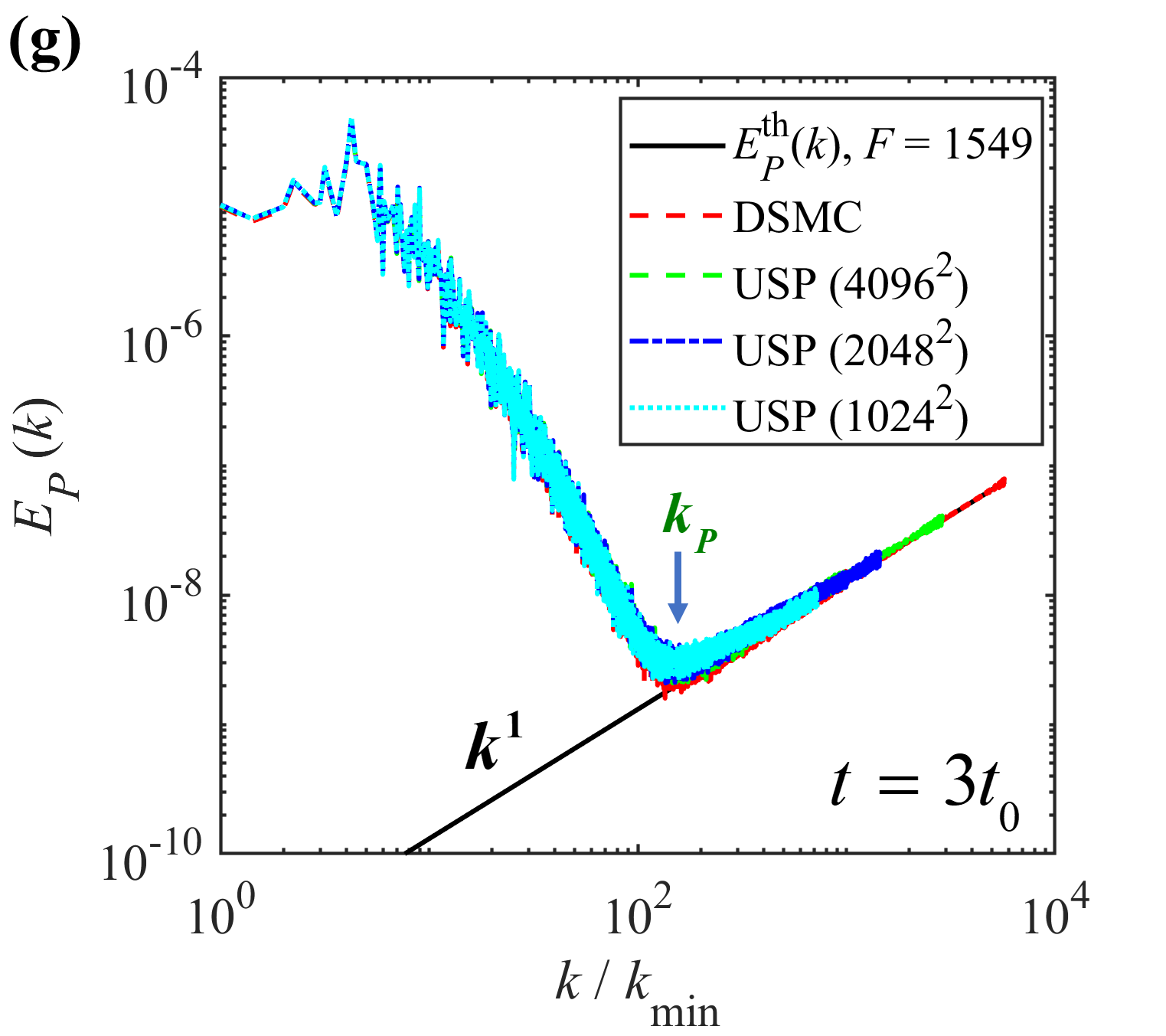}}
  \subfloat{
        \includegraphics[scale=0.42]{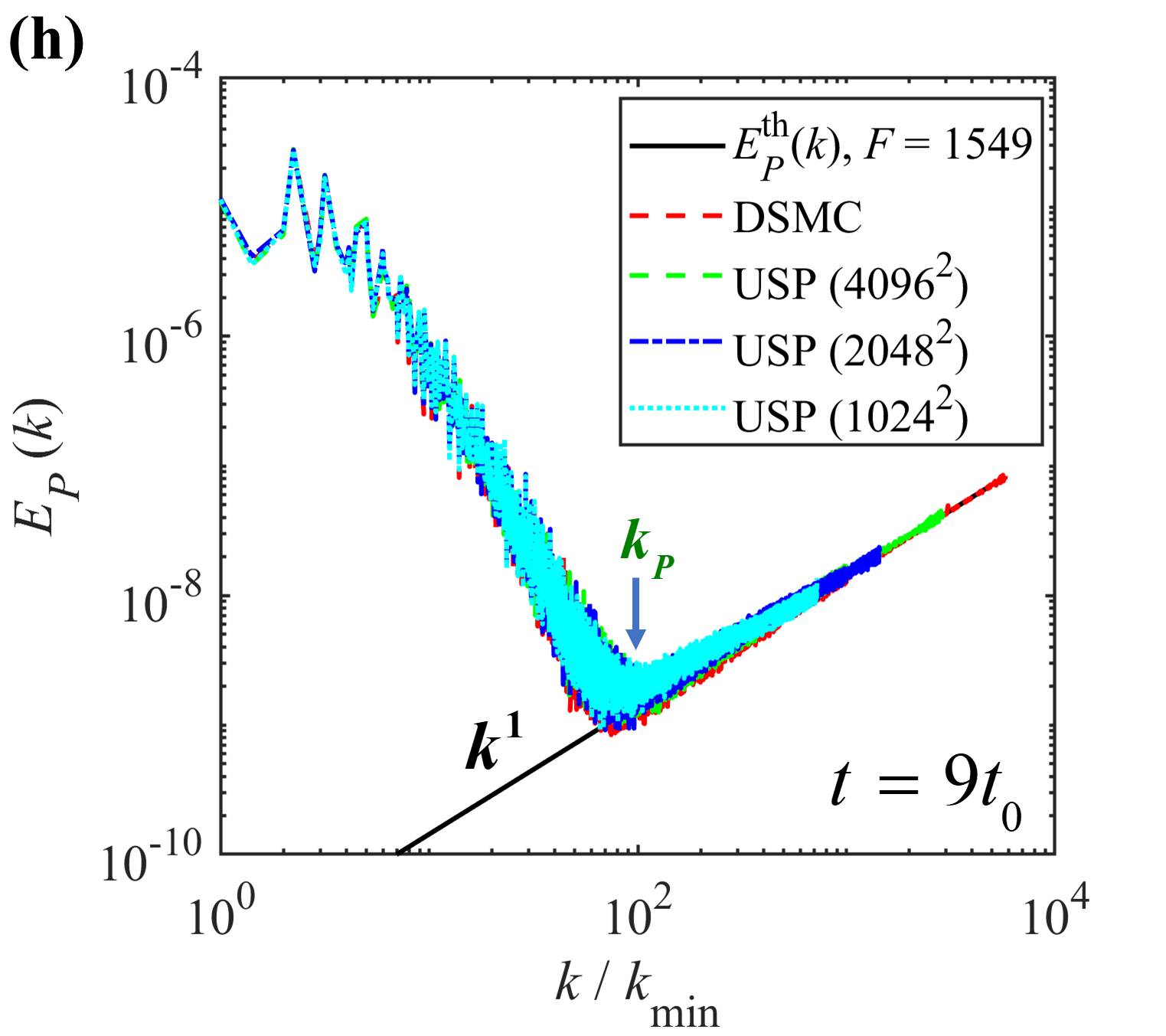}}
  \caption{\label{fig2}Energy spectra $E(k)$ and thermodynamic spectra (${{E}_{T}}(k)$, ${{E}_{n}}(k)$, ${{E}_{P}}(k)$) of 2D decaying turbulence at $t=3{{t}_{0}}$ (a, c, e, g) and $t=9{{t}_{0}}$ (b, d, f, h). The spectra of thermal fluctuations calculated from (\ref{energy_spec_eq}) and (\ref{thermal_spec}) are also displayed with $F$ = 1549.}
\end{figure}

To investigate whether the USP method can correctly reflect the effect of thermal fluctuations on turbulence, we calculate the energy spectra $E(k)$ and the thermodynamic spectra ${{E}_{g}}(k)$, where $g$ represents temperature, number density or pressure. The results for different simulation cases are shown in figure \ref{fig2}, corresponding to the time points of $t=3{{t}_{0}}$ and $t=9{{t}_{0}}$. Note that these spectra should be calculated based on the instantaneous flow field with thermal fluctuations fully preserved \citep{McMullen2022NSturbulence}. As shown in figure \ref{fig2}, the USP spectra agree well with the DSMC spectra in the low wavenumber range, suggesting that the USP method can yield consistent large-scale turbulent statistics with the DSMC method. Notably, all the USP spectra display linear growth with $k$ in the high wavenumber range, indicating the effect of thermal fluctuations. In figure \ref{fig2}, both the spectra obtained from DSMC and USP simulations align well with the theoretical spectra of thermal fluctuations, as described by (\ref{energy_spec_eq}) and (\ref{thermal_spec}) at high wavenumbers. Note that the theoretical spectra should be multiplied by the simulation ratio $F$, as the magnitude of thermal fluctuations in simulations depends on the number of simulated particles \citep{Nicolas2000DSMCcellerror,McMullen2022NSturbulence}.

We define ${{k}_{c}}$ and ${{k}_{g}}$ as the crossover wavenumbers \citep{McMullen2022NSturbulence,Bell2022thermal,Ma2023thermal} for $E(k)$ and ${{E}_{g}}(k)$, respectively. These wavenumbers give the length scales below which thermal fluctuations dominate the turbulent spectra. As shown in figure \ref{fig2}, the USP simulations at different resolutions yield identical crossover wavenumbers to those obtained by DSMC simulations. Additionally, a notable reduction in their values over time is observed when comparing the crossover wavenumbers at $t=3{{t}_{0}}$ and $t=9{{t}_{0}}$. As the crossover wavenumbers represent the length scales where turbulent fluctuations are “masked” by thermal fluctuations, it is expected that they decrease as the turbulent fluctuations continuously decay.

\begin{figure}
  \centering
  \includegraphics[scale=0.5]{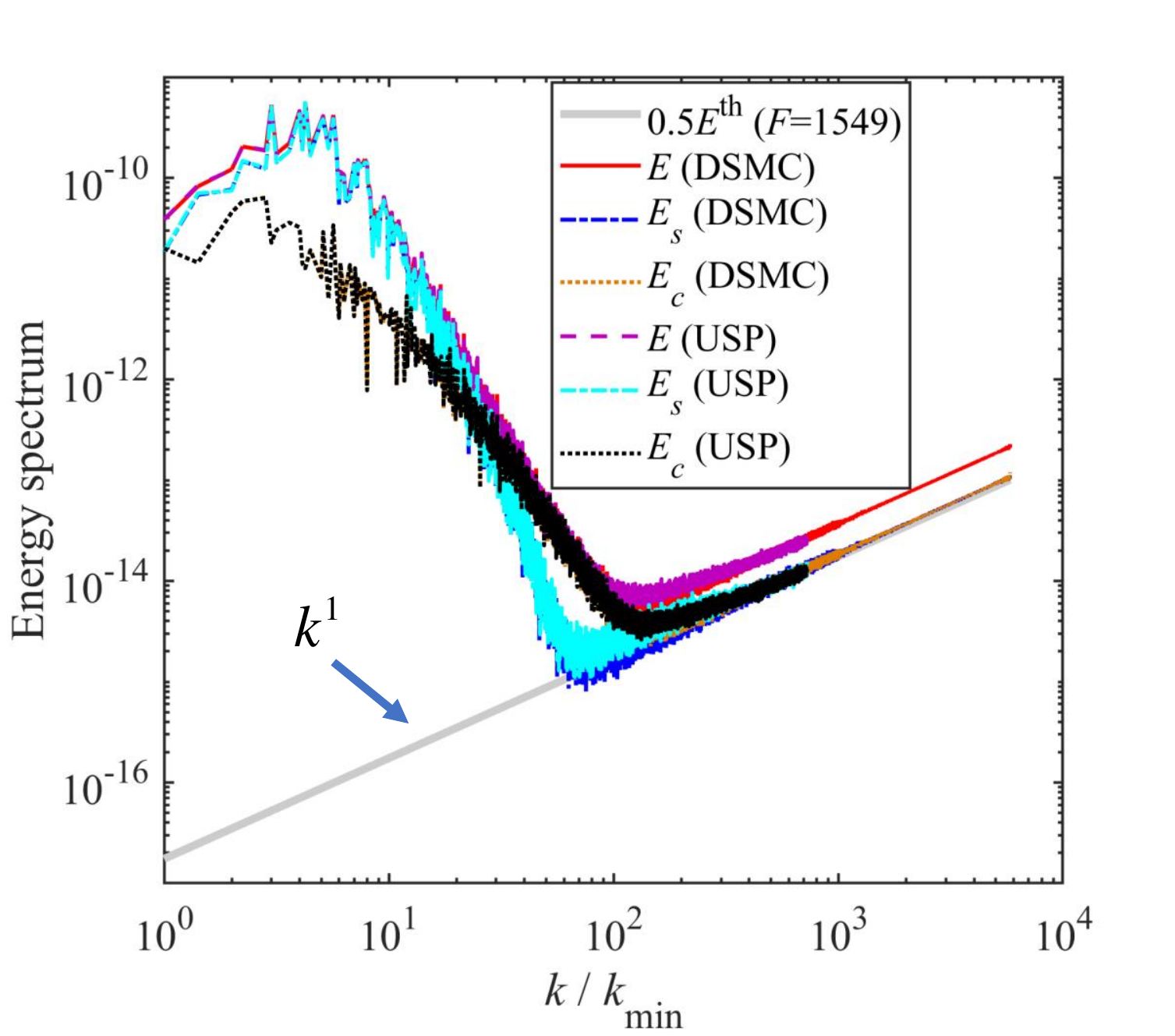}
  \caption{2D energy spectra for the velocity field and its two components obtained by DSMC and USP ($1024^2$) at $t=3{{t}_{0}}$. The theoretical spectrum of thermal fluctuations is also displayed with $F$ = 1549.}
\label{fig3}
\end{figure}

Using the Helmholtz decomposition \citep{Samtaney2001decay}, we can further investigate the effect of thermal fluctuations on the solenoidal and compressible velocity fields, respectively. Figure \ref{fig3} presents the energy spectra of the velocity field $\vec{u}$ and its two components ${{\vec{u}}_{c}}$ and ${{\vec{u}}_{s}}$ at $t=3{{t}_{0}}$. The USP results correspond to the simulation resolution of $1024^2$. As can be seen from figure \ref{fig3}, the USP spectra coincide with the DSMC spectra over the full wavenumber range. More importantly, the energy spectra of ${{\vec{u}}_{c}}$ and ${{\vec{u}}_{s}}$ overlap in the high wavenumber region, which corroborates the conclusion drawn in §\,\ref{sec:theory} that $\vec{u}_{c}^{\text{th}}$ and $\vec{u}_{s}^{\text{th}}$ satisfy the equipartition of energy in the 2D wavenumber space (see discussions before (\ref{helm_spec})). To summarize, the USP method can accurately capture the effect of thermal fluctuations on turbulence even with significantly larger time steps and cell sizes compared to the DSMC method.

\section{Three-dimensional turbulence}
\label{sec:3D}
In this section, we employ the USP method to simulate 3D decaying isotropic turbulence. The simulations begin with argon gas flows at ${{T}_{0}}=273.15\;\text{K}$ and ${{P}_{0}}=1\;\text{bar}$. The simulation domain is a cubic box with the side length of $L=2000{{\lambda }_{mic0}}$, and the periodic boundary conditions are applied in all three directions. Similar to the 2D turbulence simulations, the initial macroscopic velocity field $\vec{u}_{0}^{\text{USP}}$ is randomly generated following the relation $\vec{u}_{0}^{\text{USP}}=\vec{u}_{0}^{\text{NS}}+\vec{u}_{0}^{\text{th}}$, where $\vec{u}_{0}^{\text{NS}}$ is a divergence–free velocity field which satisfies the deterministic NS equations, and $\vec{u}_{0}^{\text{th}}$ represents the thermal fluctuations.

In this work, $\vec{u}_{0}^{\text{NS}}$ follows the special form of the energy spectrum as
\begin{equation}
  {{E}^{\text{NS}}}(k,t=0)=A{{k}^{4}}\text{exp}\left[ -2{{\left( \frac{k}{{{k}_{p}}} \right)}^{2}} \right],A=\frac{32}{3\sqrt{2\pi }}\frac{{{U}_{0}}^{2}}{{{k}_{p}}^{5}},
  \label{initial_3D_spec}
\end{equation}
where ${{k}_{p}}$ is the peak wavenumber, and ${{U}_{0}}$ is the root mean square value of $\vec{u}_{0}^{\text{NS}}$, i.e., ${{U}_{0}}={{\left\langle {{\left( \vec{u}_{0}^{\text{NS}} \right)}^{2}} \right\rangle }^{0.5}}$. In this work, we take ${{k}_{0}}=4{{k}_{\min }}$, where ${{k}_{\text{min}}}={2\pi }/{L}\;$ is the minimum wavenumber. Based on (\ref{initial_3D_spec}), the longitudinal integral length scale and the large eddy turnover time are given by \citep{Tao2020DUGKS}
\begin{equation}
  {{L}_{f0}}=\frac{3\pi }{2{{U}_{0}}^{2}}\int_{0}^{\infty }{\frac{{{E}^{\text{NS}}}(k)}{k}dk},{{T}_{e0}}=\frac{\sqrt{3}{{L}_{f}}}{{{U}_{0}}},
  \label{large_eddy}
\end{equation}
respectively. The initial dissipation rate and the Kolmogorov length scale are calculated as
\begin{equation}
  {{\varepsilon }_{0}}=2{{\nu }_{0}}\int_{0}^{\infty }{{{k}^{2}}{{E}^{\text{NS}}}(k)dk},{{\eta }_{0}}={{\left( {{{\nu }_{0}}^{3}}/{{{\varepsilon }_{0}}}\; \right)}^{{1}/{4}\;}},
\end{equation}
respectively. The initial turbulent Mach number ${{M}_{t0}}$ is calculated using (\ref{Mach_Re_2D}), and the initial Taylor microscale ${{\lambda }_{0}}$ and the corresponding Reynolds number are given by \citep{Pope2000}
\begin{equation}
  {{\lambda }_{0}}=\sqrt{\frac{5{{\nu }_{0}}{{U}_{0}}^{2}}{{{\varepsilon }_{0}}}},R{{e}_{\lambda 0}}=\frac{{{U}_{0}}{{\lambda }_{0}}}{\sqrt{3}{{\nu }_{0}}},
  \label{Re_lamda3D}
\end{equation}
respectively.

Table \ref{3Dparameter} shows the parameters of USP simulations, where ${{M}_{t0}}$ ranges from 0.6 to 0.9, and $R{{e}_{\lambda 0}}$ increases with ${{M}_{t0}}$. Based on the discussions in §\,\ref{sec:2D}, the USP simulations are performed with larger time steps and cell sizes compared to those typically used in DSMC simulations. The average number of simulated particles per cell is 100, resulting in a total of 13.42 billion particles, each of which represents 1838 real molecules.
\begin{table}
  \centering
    \begin{tabular}{*{8}{c}}
    %\hline
    Resolution (${{N}_{c}}^{3}$) & $R{{e}_{\lambda 0}}$ &  ${{M}_{t0}}$  & ${{{L}_{f0}}}/{{{\eta }_{0}}}\;$ & $\left\langle {{N}_{p}} \right\rangle$ & ${\Delta t}/{{{\tau }_{mic0}}}\;$ &  ${\Delta {{L}_{cell}}}/{{{\lambda }_{mic0}}}\;$ & ${{k}_{\max }}{{\eta}_{0}}$ \\
    \midrule
    ${{512}^{3}}$	 &68.8	  &0.6	  &20.5	  &100	 &1  	  &3.9    &7.84  \\
    ${{512}^{3}}$	 &86.1	  &0.75	  &22.9	  &100	 &0.8 	&3.9    &7.01  \\
    ${{512}^{3}}$	 &103.3	  &0.9	  &25.1	  &100	 &0.8 	&3.9    &6.40  \\\hline
    \end{tabular}
    \caption{USP simulation parameters for 3D decaying isotropic turbulence. All the simulations are performed with the initial conditions of ${{T}_{0}}=273.15\;\text{K}$ and ${{P}_{0}}=1\;\text{bar}$.}
  \label{3Dparameter}
\end{table}

In addition to the USP simulations, we numerically solved the 3D deterministic compressible NS equations using the direct numerical simulation (DNS) method \citep{Pope2000}. The effect of thermal fluctuations on turbulence can then be analyzed by comparing the USP and DNS results. Specifically, three DNS cases are performed with the same $R{{e}_{\lambda 0}}$ and ${{M}_{t0}}$ as USP simulations. The gas thermodynamic properties in DNS are identical to those in USP simulations. The initial values of $\rho$, $T$, and $P$ are uniformly set within the DNS domain, and the initial velocity field is directly obtained from $\vec{u}_{0}^{\text{NS}}$ generated during the USP initialization procedures. For the numerical scheme of DNS, considering that ${{M}_{t0}}$ is high, we utilize a hybrid scheme proposed by \citet{Wang2010hybrid}, which combines an eighth-order compact central finite difference scheme \citep{Lele1992compact} for smooth regions and a seventh-order weighted essentially non-oscillatory (WENO) scheme \citep{Shu2000WENO} for shock regions. The time steps for all the DNS cases are smaller than 0.001${{T}_{e0}}$, and the space resolutions in DNS are the same as those in USP simulations (see table \ref{3Dparameter}). Previous grid refinement studies of the hybrid scheme \citep{WangJC2011shocklets,Wang2012compressibilityJFM} have shown that the resolution parameters ${{k}_{\max }}{{\eta }_{0}}\ge 3.3$ are enough for the convergence of turbulent statistics.
\subsection{Basic turbulent statistics}
\label{subsec:turbstats}
\begin{figure}
  \centering
  \subfloat{
      \includegraphics[scale=0.5]{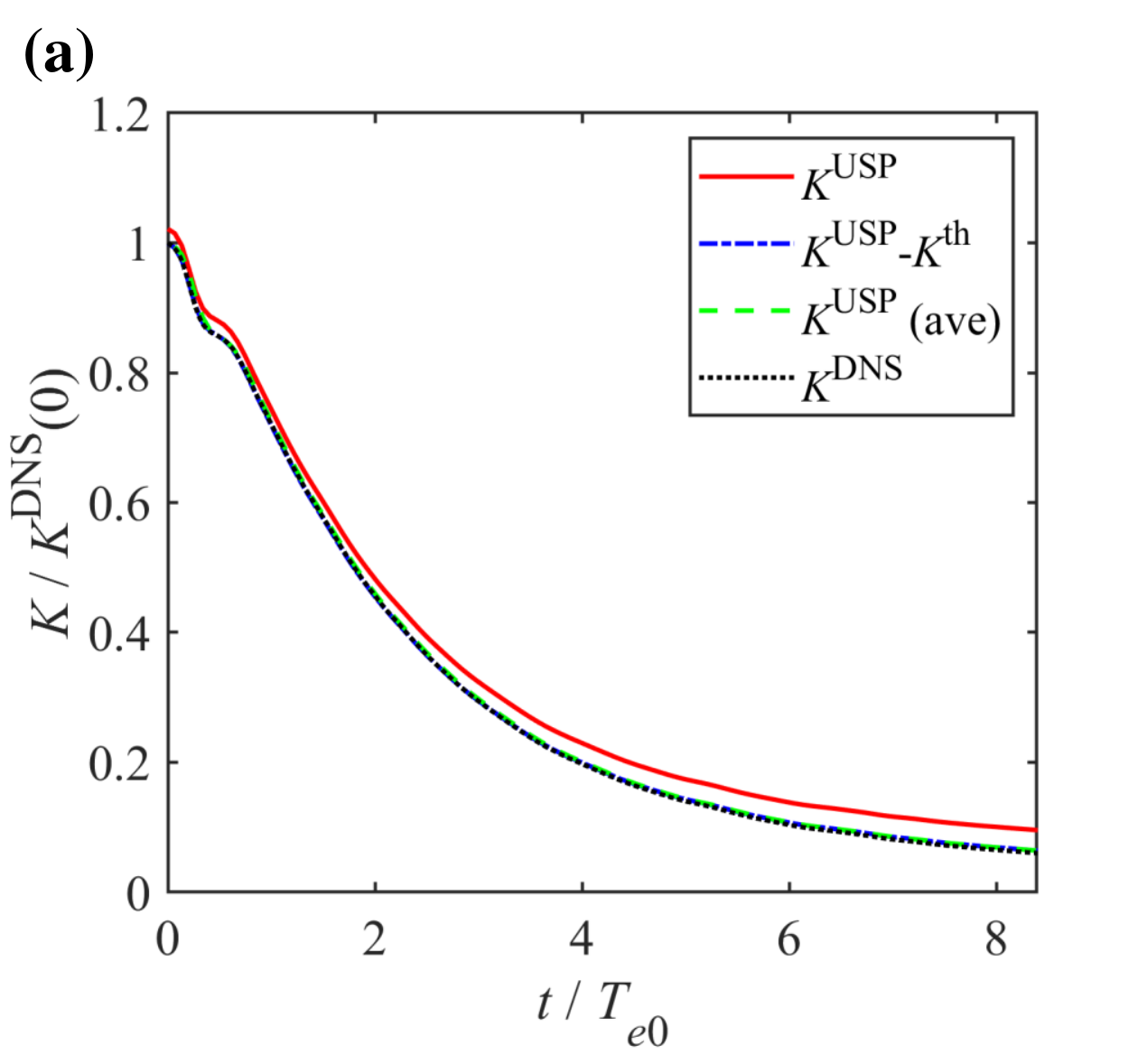}}
  \subfloat{
      \includegraphics[scale=0.5]{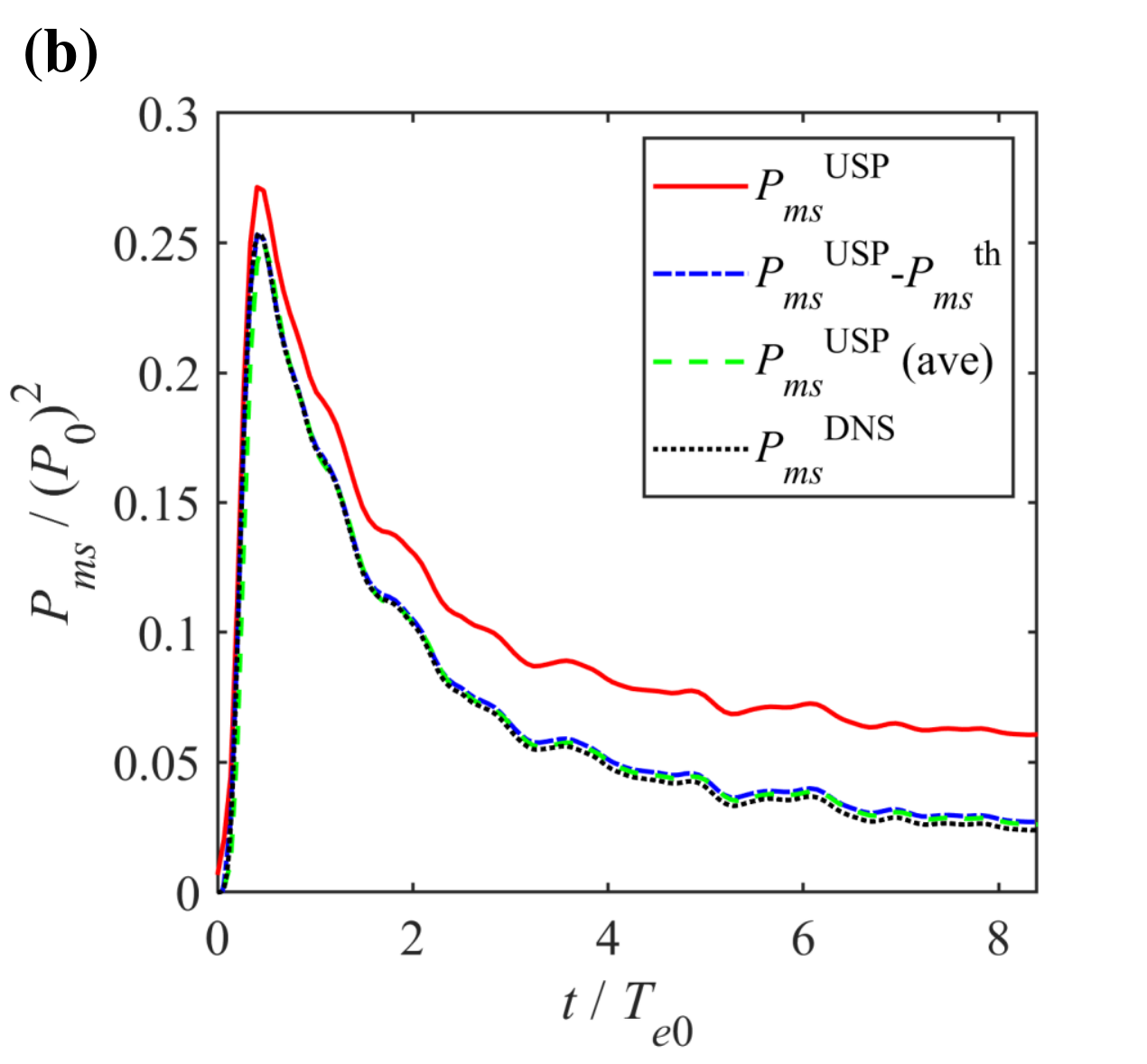}}\\
  \caption{\label{fig4}Time evolution of (a) normalized turbulent kinetic energy, and (b) normalized mean square value of pressure fluctuations, for the case with $R{{e}_{\lambda 0}}$= 103.3 and ${{M}_{t0}}$= 0.9.}
\end{figure}

In previous studies employing the DSMC method \citep{McMullen2022NSturbulence,McMullen2023thermal}, researchers have demonstrated the statistical independence between thermal fluctuations and turbulent fluctuations predicted by the deterministic NS equations. Consequently, it is anticipated that the mean square fluctuations in USP simulations at a given time can be expressed as $\left\langle {{\left( \delta {{a}^{\text{USP}}} \right)}^{2}} \right\rangle =\left\langle {{\left( \delta {{a}^{\text{DNS}}} \right)}^{2}} \right\rangle +\left\langle {{\left( \delta {{a}^{\text{th}}} \right)}^{2}} \right\rangle $, where $\delta {{a}^{\text{DNS}}}$ represents the turbulent fluctuations predicted by the DNS method. To illustrate this relation, figure \ref{fig4} presents the simulation results for the turbulent kinetic energy $K$ and the mean square pressure fluctuations $P_{ms}$ in the case of $R{{e}_{\lambda 0}}$= 103.3 and ${{M}_{t0}}$= 0.9. $K$ is normalized by the initial value of DNS, and $P_{ms}$ is normalized by the square of initial pressure. As observed in figure \ref{fig4}, the results obtained from USP simulations (solid lines) are notably larger than those obtained from DNS (dotted lines), indicating the presence of thermal fluctuations, and the corresponding results (${{K}^{\text{th}}}$ and $P_{ms}^{\text{th}}$) can be obtained at each time point using (\ref{eq_kturb}) and (\ref{eq_Press}) with $F=1838$. By subtracting ${{K}^{\text{th}}}$ and $P_{ms}^{\text{th}}$ directly from the USP results, we obtain results (dash-dotted lines) that perfectly align with those obtained from DNS. Note that the effects of thermal fluctuations can also be reduced by averaging the USP flow field over time intervals that are long compared to the simulation time step $\Delta t$ but short compared to the Kolmogorov time scale ${{\tau }_{\eta }}$ \citep{Gallis2021turbulence}. The USP results obtained after this short-time average procedure are also shown in figure \ref{fig4} (dashed lines), which are in good agreement with the DNS results.

\begin{figure}
  \centering
  \subfloat{
      \includegraphics[scale=0.63]{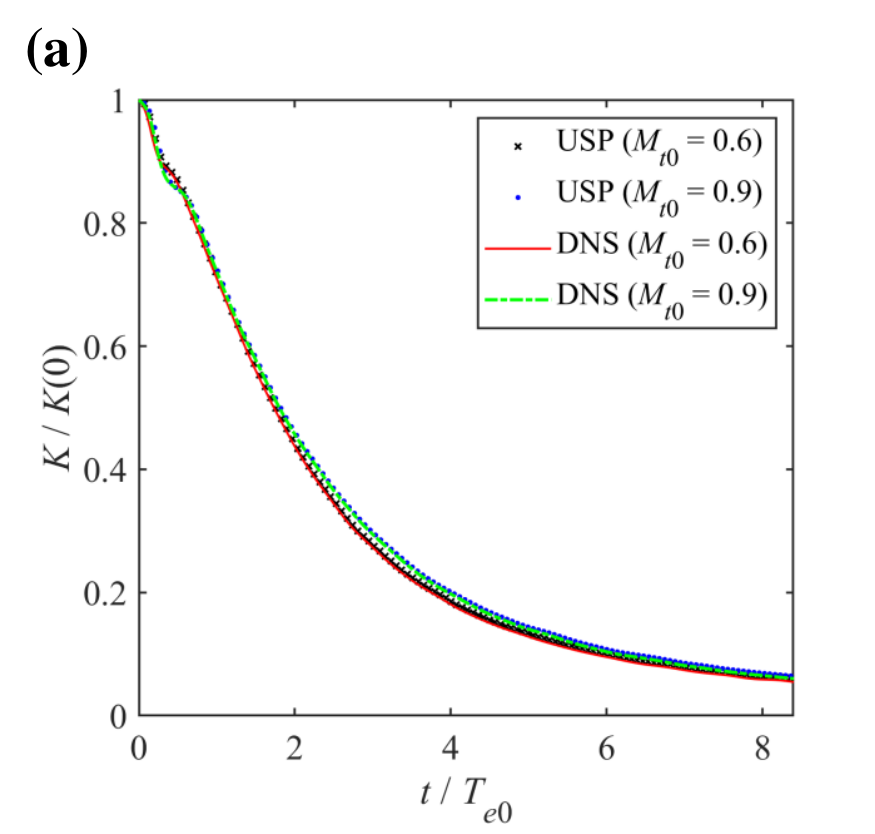}}
  \subfloat{
      \includegraphics[scale=0.63]{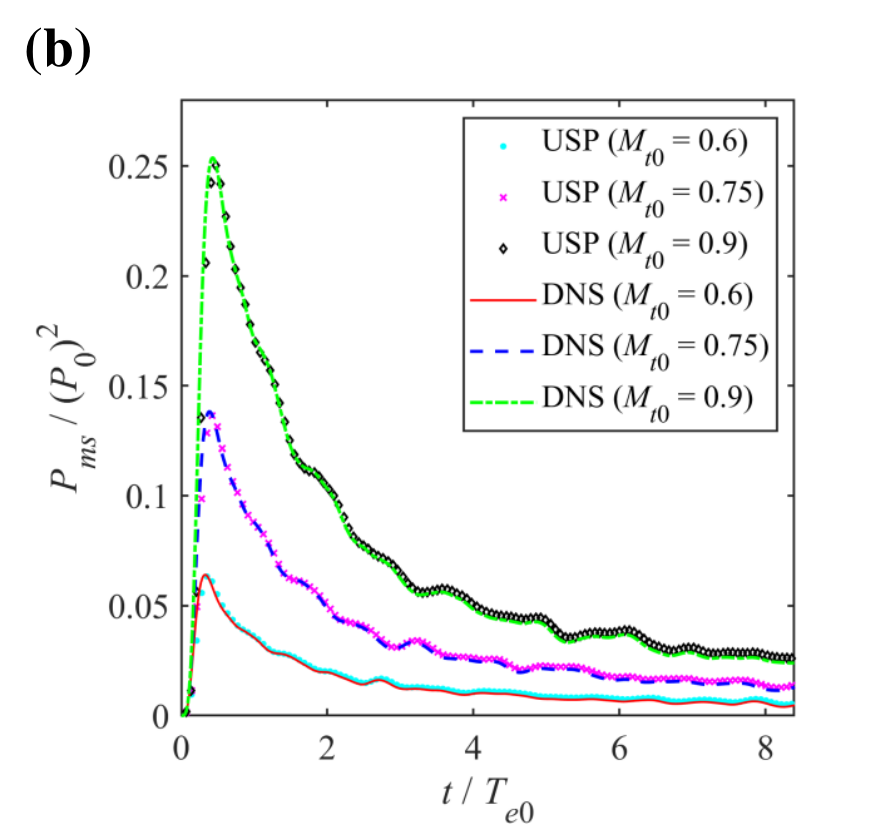}}\\
  \vspace{-1em}
  \subfloat{
      \includegraphics[scale=0.63]{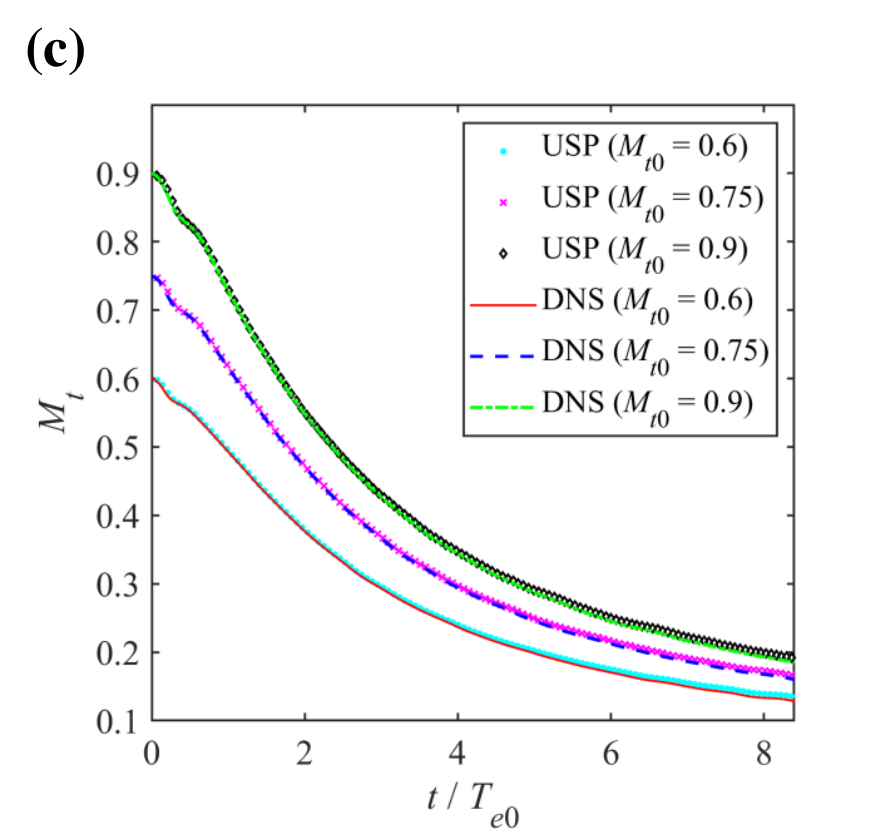}}
  \subfloat{
      \includegraphics[scale=0.63]{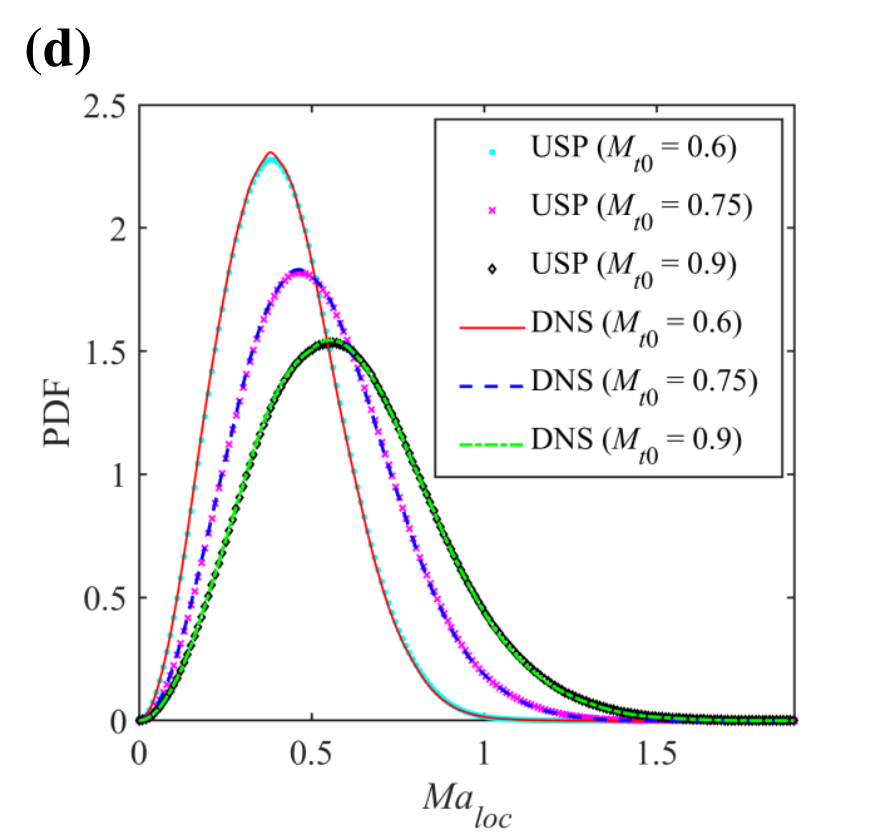}}\\
  \caption{\label{fig5}Panels (a)–(c) display the time evolutions of normalized turbulent kinetic energy, normalized mean square pressure fluctuations and turbulent Mach number, respectively. Panel (d) displays the PDFs of the local Mach number at $t=1.4{{T}_{e0}}$. All the USP results are obtained based on the short-time average flow field without thermal fluctuations.}
\end{figure}

Figures \ref{fig5}(a)–(c) show the temporal evolutions of ${K}/{K(0)}$, ${{{P}_{ms}}}/{\left({{P}_{0}}\right)^2}$ and $M_t$ obtained from both USP and DNS simulations for cases with different ${{M}_{t0}}$. The USP results correspond to the short-time average flow field. In figure \ref{fig5}(a), since the time histories of ${K}/{K(0)}$ for different ${{M}_{t0}}$ almost overlap, only the results for ${{M}_{t0}}=0.6$ and ${{M}_{t0}}=0.9$ are presented. As observed from figures \ref{fig5}(a)–(c), the USP results exhibit excellent agreement with the DNS results throughout the entire time range. For cases with higher ${{M}_{t0}}$, the fluctuations of thermodynamic variables are amplified to greater magnitudes due to the increase of compressibility. To further validate the accuracy of USP simulations, we compare the probability density functions (PDFs) of the local Mach number $M{{a}_{loc}}$ obtained from both the USP and DNS simulations. $M{{a}_{loc}}$ is defined as \citep{Samtaney2001decay,Tao2020DUGKS}
\begin{equation}
  M{{a}_{loc}}=\frac{{{\left( \vec{u}\cdot \vec{u} \right)}^{{1}/{2}\;}}}{\sqrt{\gamma RT}}.
\end{equation}
Figure \ref{fig5}(d) shows the PDFs at $t=1.4{{T}_{e0}}$, where the USP results exhibit good agreement with the DNS results across all three cases.

\subsection{Effect of thermal fluctuations on spectra}
\label{subsec:spectra}
\begin{figure}
  \centering
  \includegraphics[scale=0.55]{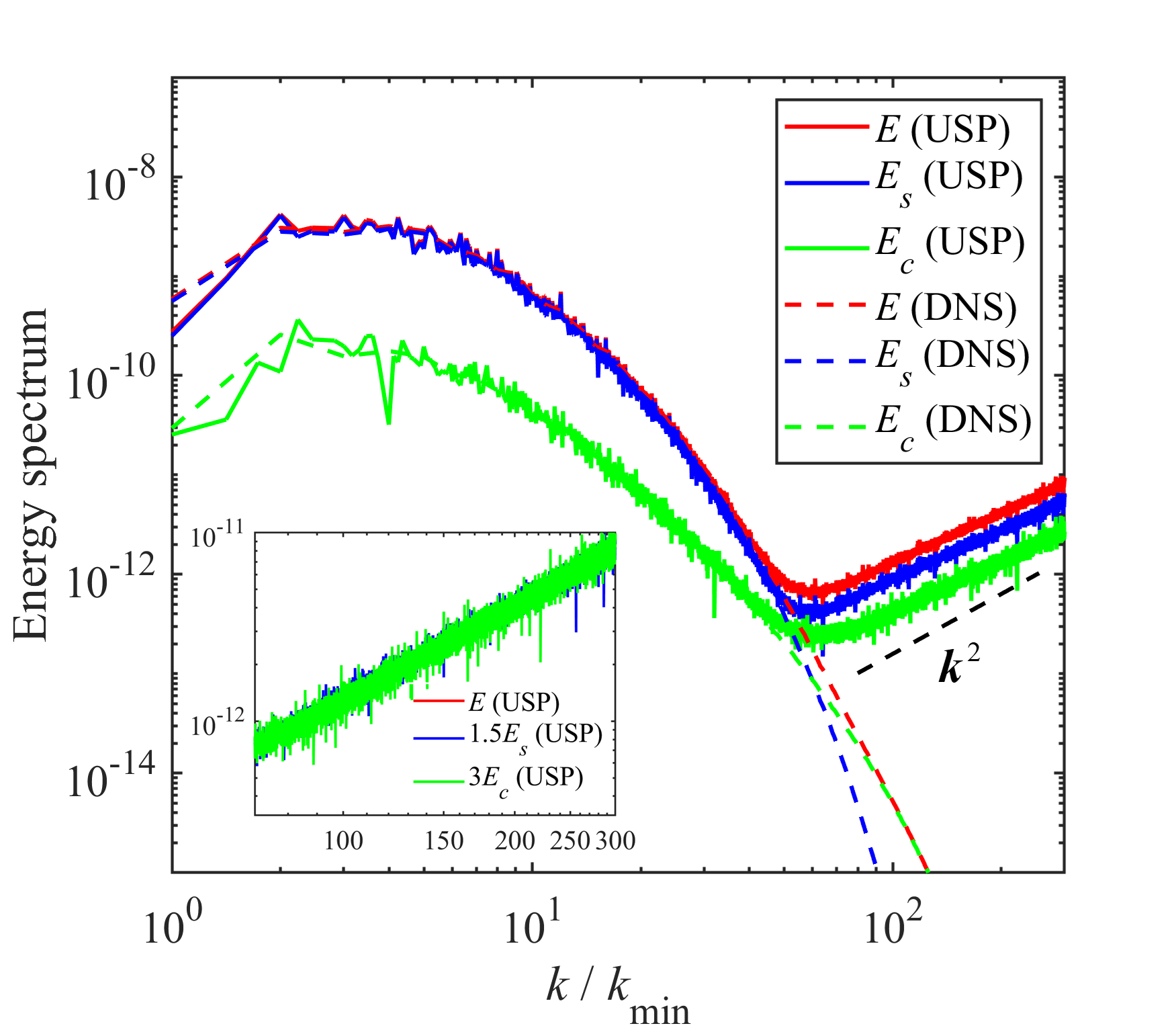}
  \caption{Energy spectra for the velocity field and its two components obtained from USP and DNS simulations at $t=1.4{{T}_{e0}}$, for the case of $R{{e}_{\lambda 0}}$= 103.3 and ${{M}_{t0}}$= 0.9. The inset shows the relationship between $E(k)$, ${{E}_{c}}(k)$ and ${{E}_{s}}(k)$ obtained from USP simulations at large wavenumbers.}
\label{fig6}
\end{figure}

In previous relevant numerical studies on 3D turbulence \citep{Bell2022thermal,McMullen2022NSturbulence,McMullen2023thermal}, researchers mainly focused on the effect of thermal fluctuations on the spectra of the velocity field $\vec{u}$. By employing the Helmholtz decomposition, we can further consider the effect of thermal fluctuations on spectra of the solenoidal and compressible velocity components (i.e., ${{\vec{u}}_{s}}$ and ${{\vec{u}}_{c}}$). Figure \ref{fig6} presents the USP and DNS spectra of $\vec{u}$, ${{\vec{u}}_{c}}$ and ${{\vec{u}}_{s}}$ at $t=1.4{{T}_{e0}}$, for the case of ${{M}_{t0}}=0.9$. The spectra are plotted against the dimensionless wavenumber ${k}/{{{k}_{\min}}}$. Except for being slightly noisy, the USP spectra agree well with the DNS spectra at small wavenumbers. In the high wavenumber range, the DNS spectra exhibit a continuous decrease, whereas the USP spectra exhibit a quadratic growth with respect to $k$, which corresponds to the effect of thermal fluctuations \citep{Bell2022thermal,McMullen2022NSturbulence,Bandak2022thermalPRE}. In figure \ref{fig6}, we further compare $E(k)$, ${{E}_{c}}(k)$ and ${{E}_{s}}(k)$ obtained by USP at large wavenumbers (see the inset), illustrating the relation of $E(k)=1.5{{E}_{s}}(k)=3{{E}_{c}}(k)$. The USP results support the conclusion drawn in §\,\ref{sec:theory} that the energy of $\vec{u}_{s}^{\text{th}}$ is twice the energy of $\vec{u}_{c}^{\text{th}}$ in the wavenumber space (see discussions before (\ref{helm_spec})).

\begin{figure}
  \centering
  \subfloat{
      \includegraphics[scale=0.5]{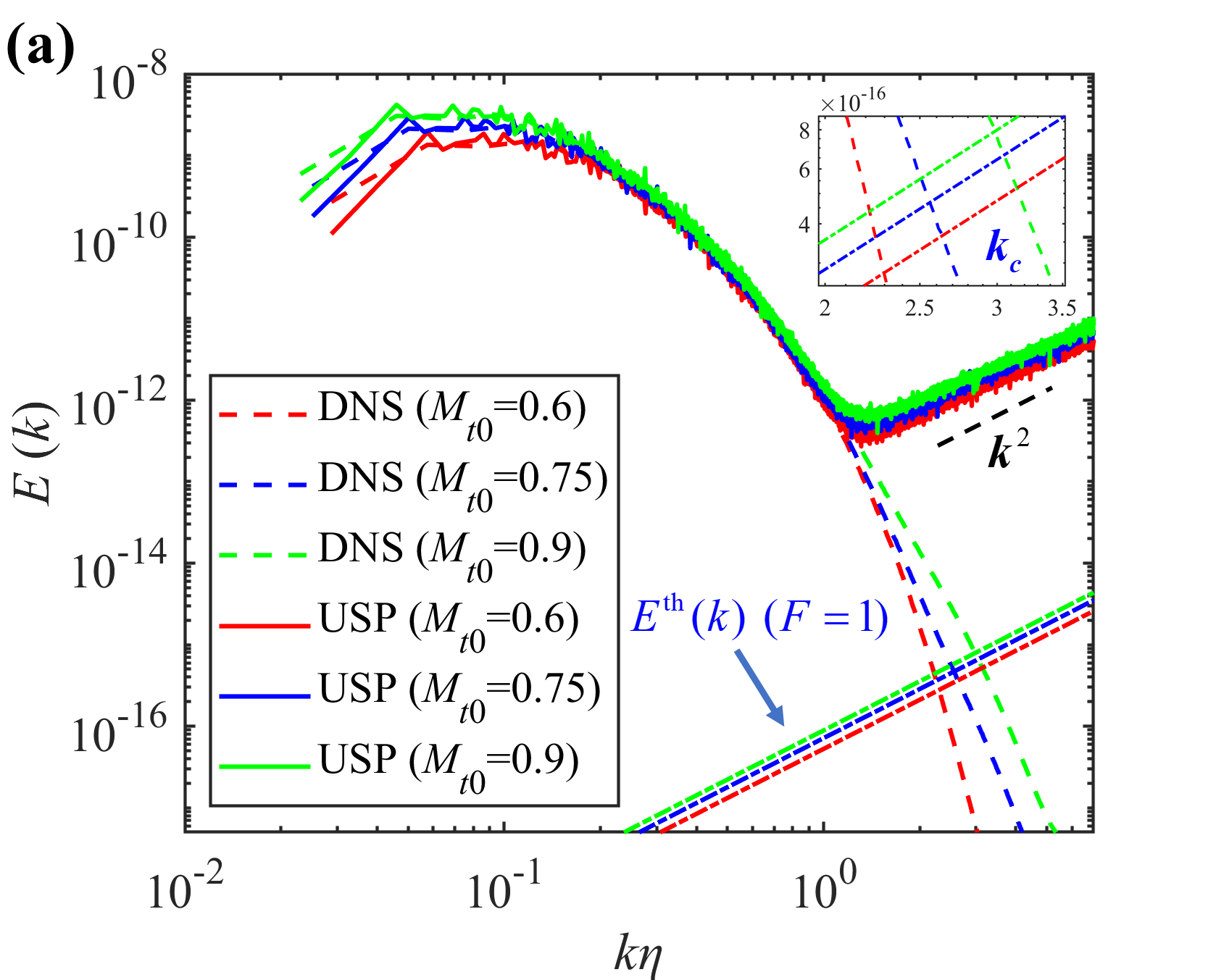}}
  \subfloat{
      \includegraphics[scale=0.5]{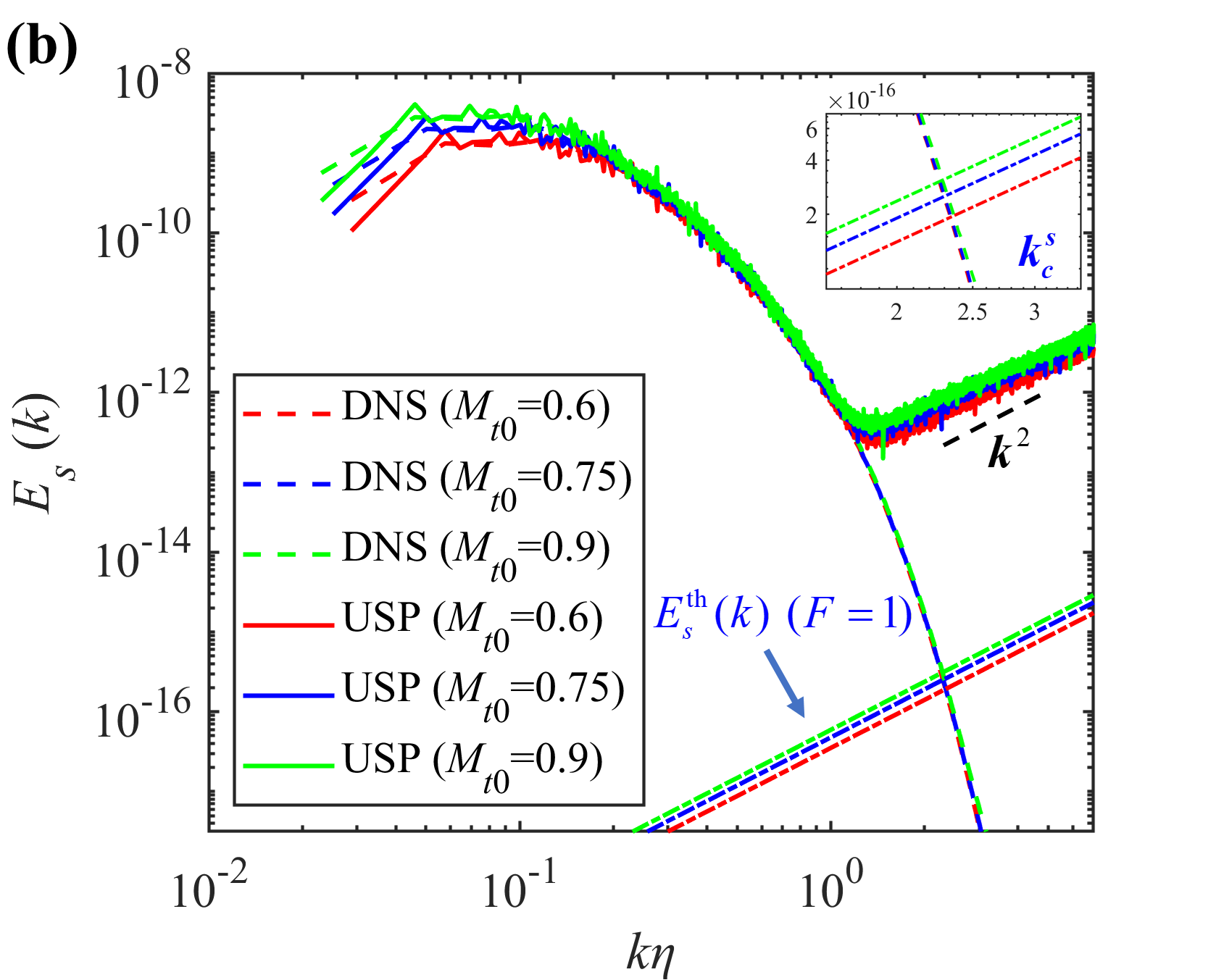}}\\
  \vspace{-1em}
  \subfloat{
      \includegraphics[scale=0.5]{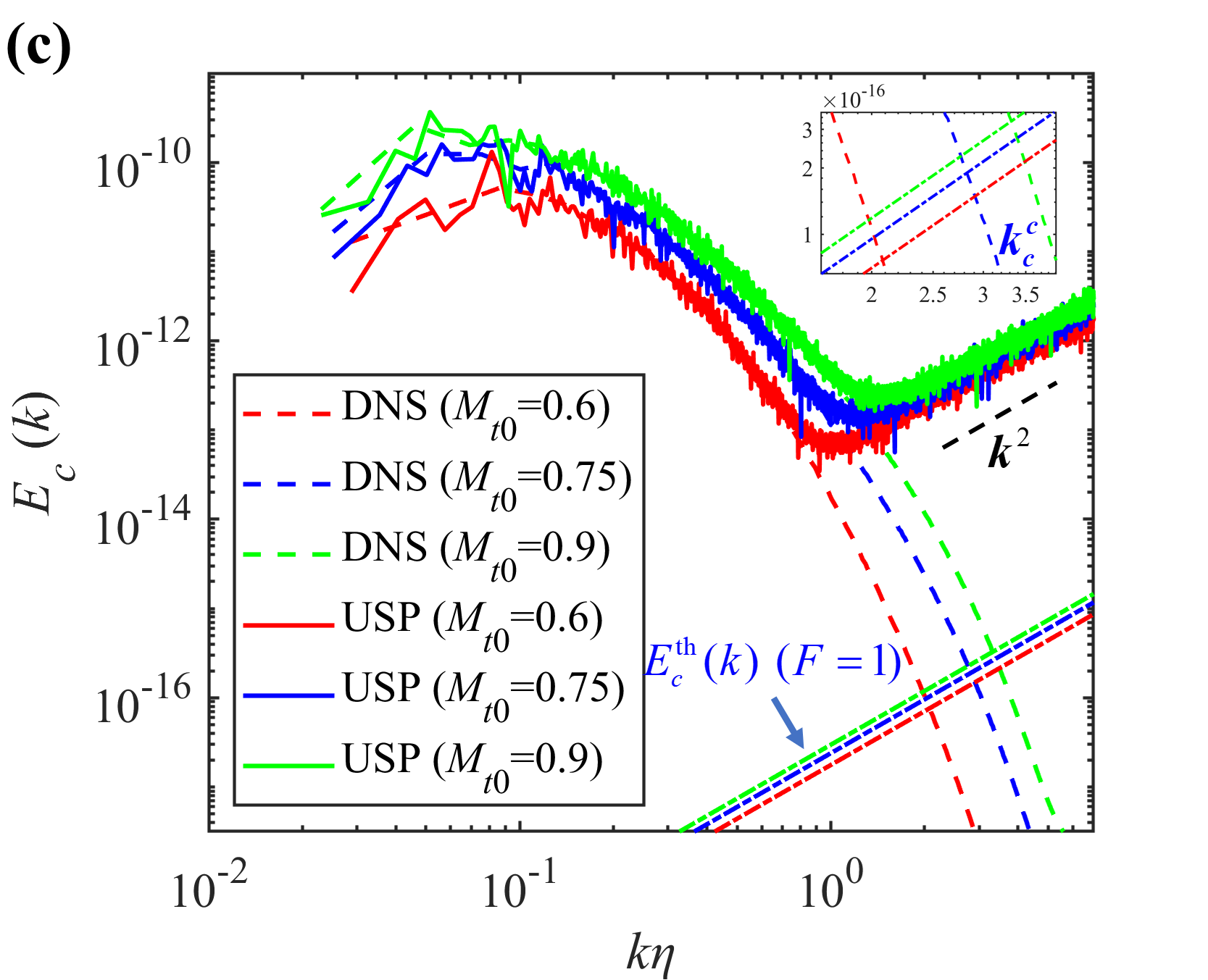}}
  \caption{\label{fig7}$E(k)$ (a), ${{E}_{s}}(k)$ (b) and ${{E}_{c}}(k)$ (c) obtained from USP and DNS simulations at $t=1.4{{T}_{e0}}$, for cases with different ${{M}_{t0}}$. The thermal fluctuation spectra with $F$ = 1 are also shown for comparison. The insets are the enlarged views showing the intersection points between thermal fluctuation spectra and DNS spectra (i.e., the crossover wavenumbers for the real gases).}
\end{figure}

In figure \ref{fig7}, we present $E(k)$, ${{E}_{s}}(k)$ and ${{E}_{c}}(k)$ obtained from USP and DNS simulations at $t=1.4{{T}_{e0}}$ under different ${{M}_{t0}}$ conditions, in order to study the effect of compressibility on spectra. The spectra are plotted against the dimensionless wavenumber $k\eta$, where $\eta$ is the Kolmogorov length scale obtained by DNS. The crossover wavenumbers ${{k}_{c}}$, $k_{c}^{s}$ and $k_{c}^{c}$ are estimated as the intersections of $E(k)$, ${{E}_{s}}(k)$ and ${{E}_{c}}(k)$ obtained by DNS with the thermal fluctuation spectra \citep{McMullen2022NSturbulence}.

According to the discussions in §\,\ref{sec:2D}, the thermal fluctuation spectra obtained by USP simulations are overestimated due to the use of a simulation ratio $F>\text{1}$. By setting $F$ = 1, we can obtain the spectra of thermal fluctuations corresponding to the real gases. As observed from figure \ref{fig7}(a), ${{k}_{c}}\eta$ lies between 2.3 and 3 for $F = 1$, corresponding to ${\eta }/{{{l}_{c}}}$ in the range of 0.37 to 0.48 (${{l}_{c}}={2\pi }/{{{k}_{c}}}$), indicating that thermal fluctuations dominate $E(k)$ at spatial scales slightly larger than the Kolmogorov length scale. The similar results were also reported by \citet{McMullen2022NSturbulence} and \citet{Bell2022thermal} in their simulations of 3D turbulence, where they found ${\eta }/{{{l}_{c}}}\approx 0.5$. More interestingly, with the increase of ${{M}_{t}}$, it is noteworthy that $k_{c}^{s}\eta$ remains relatively stable around 2.3 (see figure \ref{fig7}(b)), whereas $k_{c}^{c}\eta $ changes significantly from 2.1 to 3.3 (see figure \ref{fig7}(c)). This observation suggests that the influence of thermal fluctuations on ${{\vec{u}}_{c}}$ is more responsive to changes in compressibility compared to that on ${{\vec{u}}_{s}}$. Despite the USP simulations being performed with a large simulation ratio ($F$ = 1838), the trends of the crossover wavenumbers predicted by USP are completely consistent with those observed in real gases.

\begin{figure}
  \centering
  \subfloat{
      \includegraphics[scale=0.5]{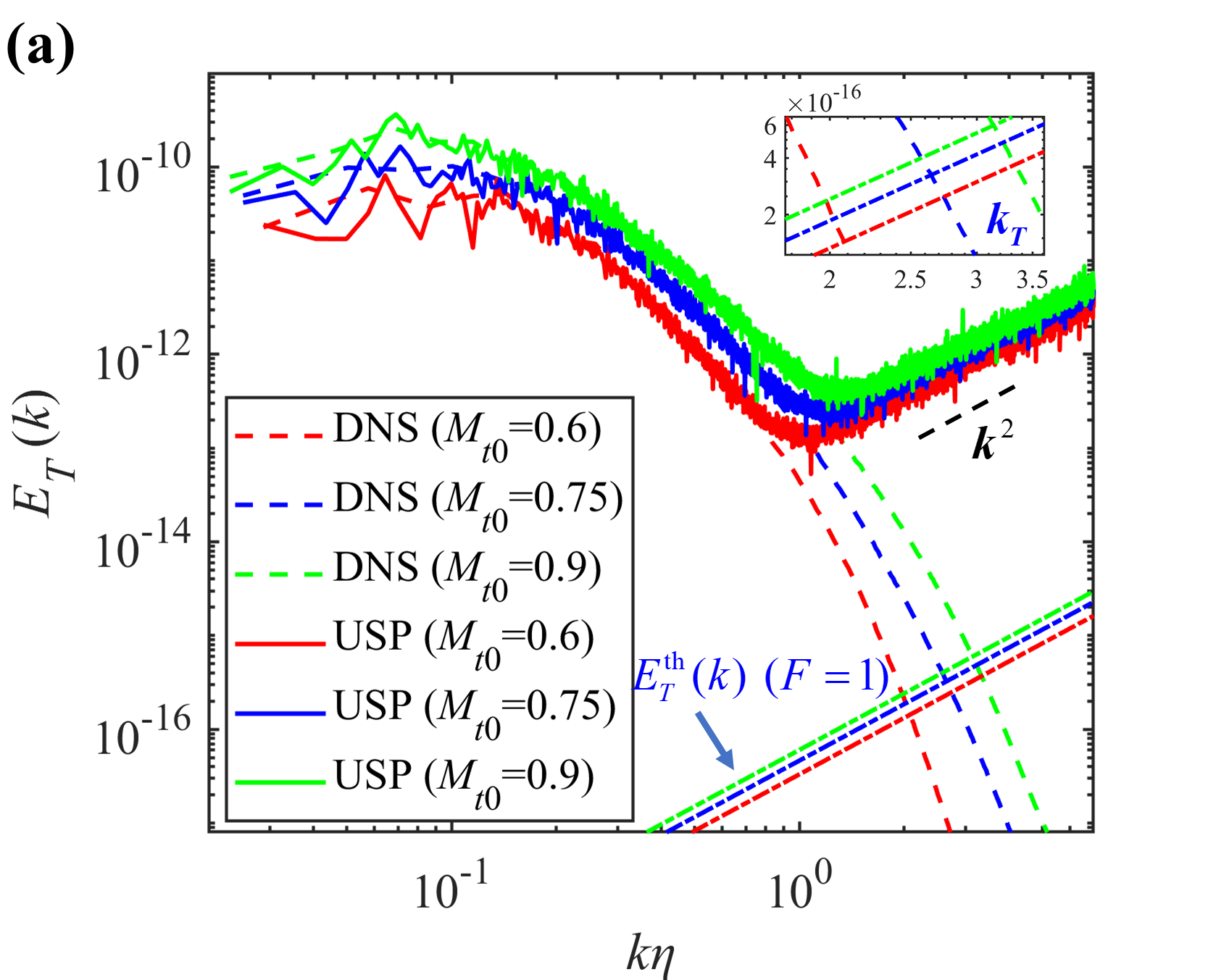}}
  \subfloat{
      \includegraphics[scale=0.5]{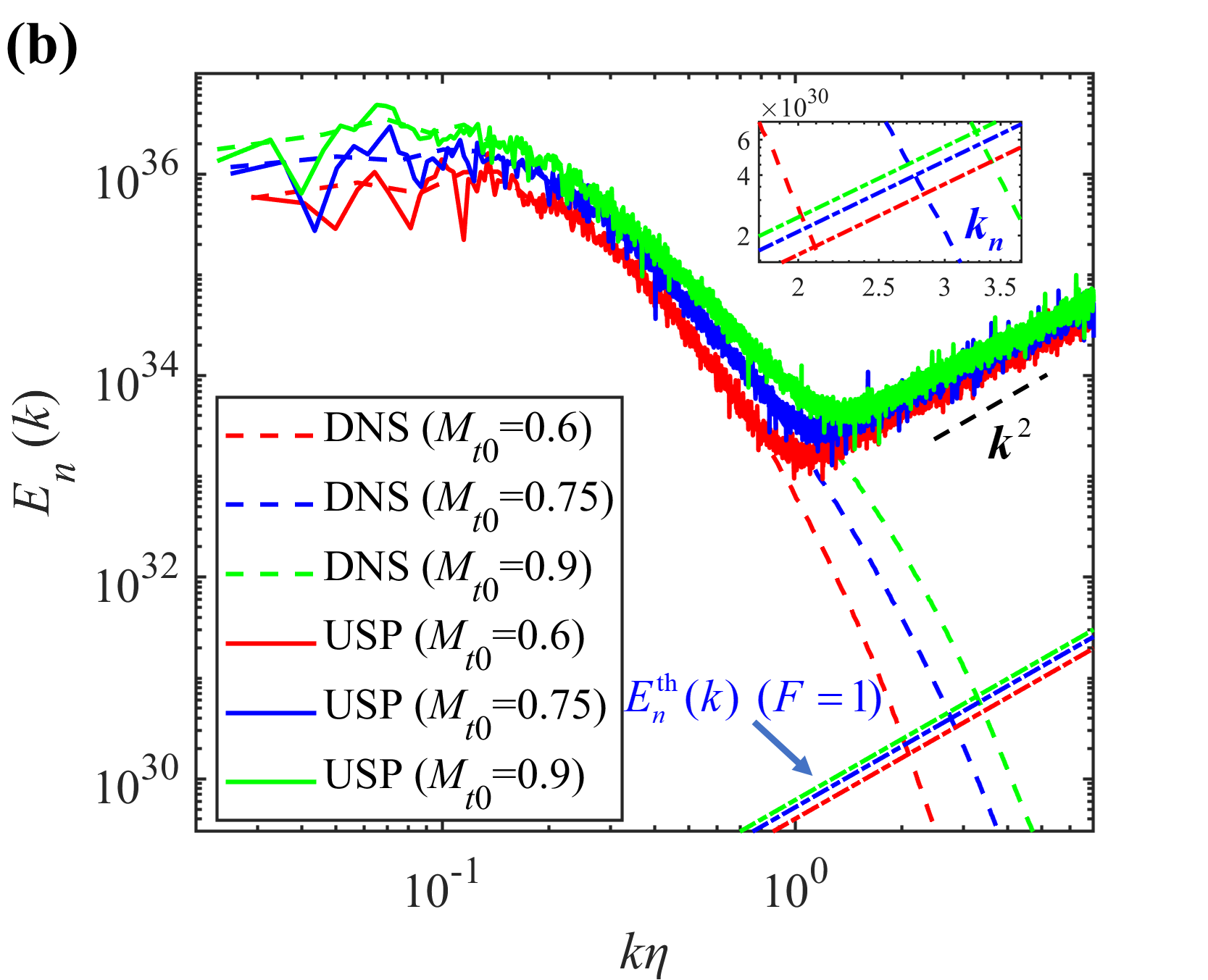}}\\
  \vspace{-1em}
  \subfloat{
      \includegraphics[scale=0.5]{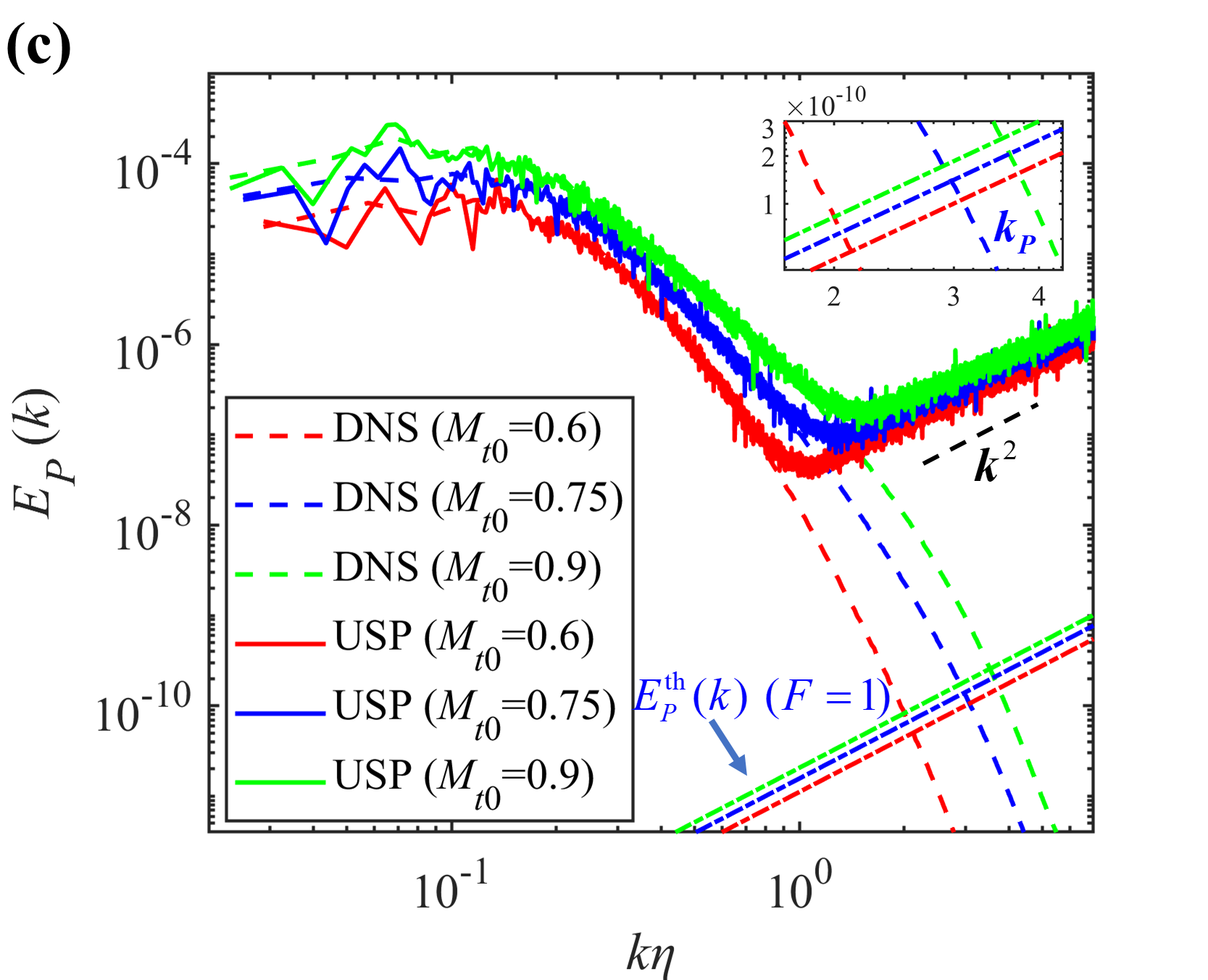}}
  \caption{\label{fig8}${{E}_{T}}(k)$ (a), ${{E}_{n}}(k)$ (b) and ${{E}_{P}}(k)$ (c) obtained from USP and DNS simulations at $t=1.4{{T}_{e0}}$, for cases with different ${{M}_{t0}}$. The thermal fluctuation spectra with $F$ = 1 are also shown for comparison. The insets are the enlarged views showing the intersection points between thermal fluctuation spectra and DNS spectra (i.e., the crossover wavenumbers for the real gases).}
\end{figure}

\begin{figure}
  \centering
  \subfloat{
      \includegraphics[scale=0.5]{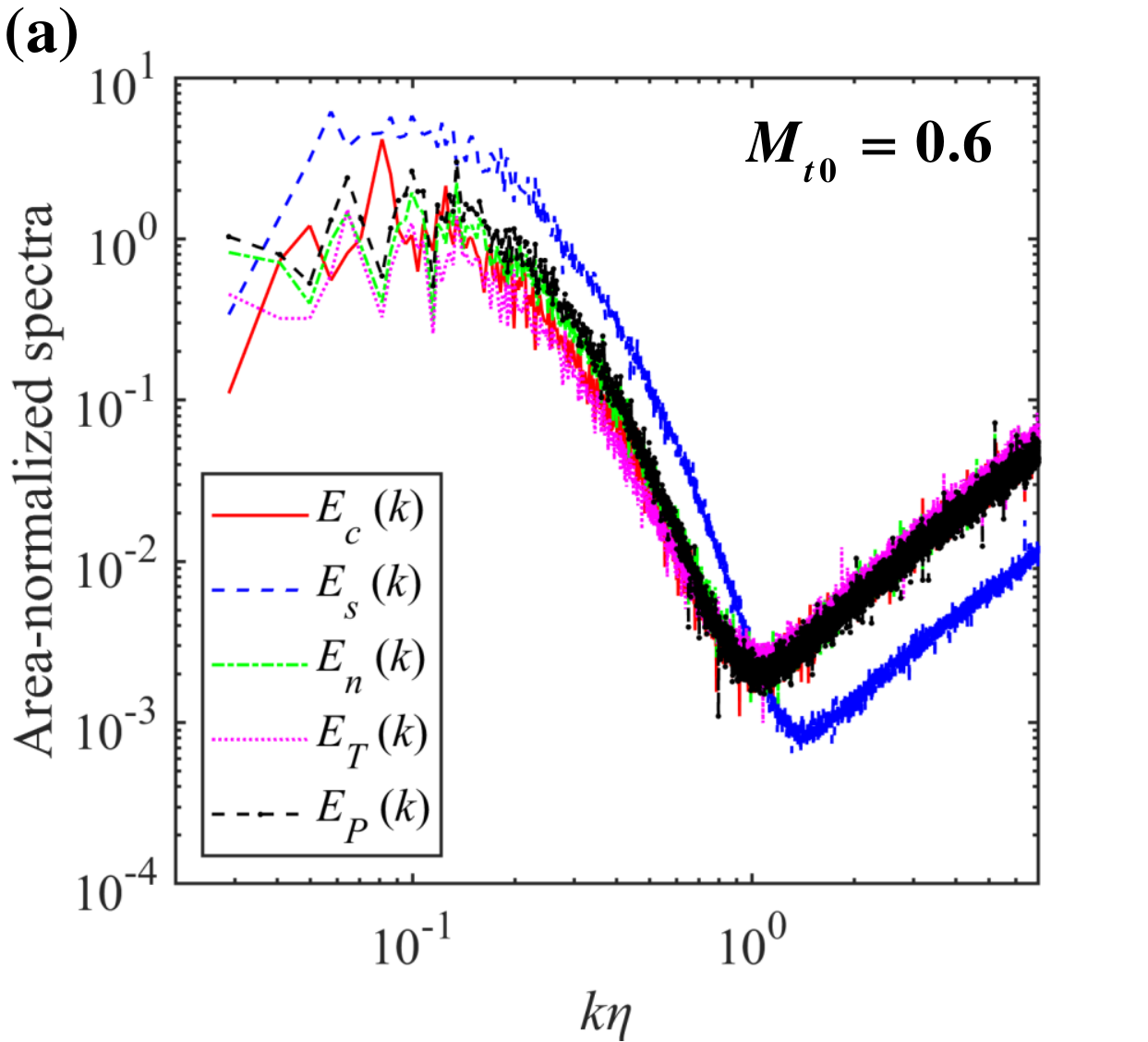}}
  \subfloat{
      \includegraphics[scale=0.5]{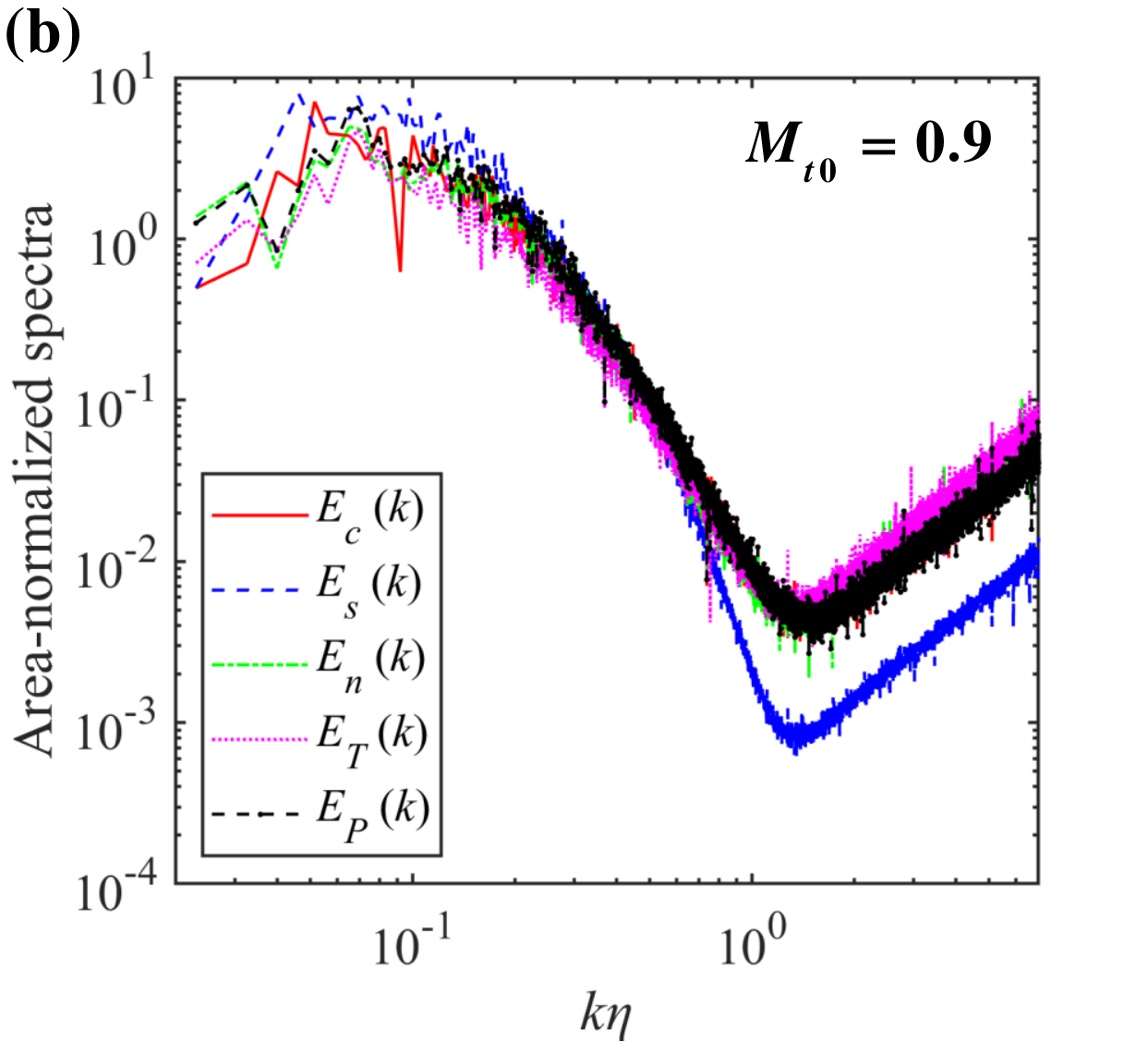}}
  \caption{\label{fig9}Normalized USP spectra of compressible velocity component, solenoidal velocity component, number density, temperature and pressure at $t=1.4{{T}_{e0}}$, for cases with ${{M}_{t0}}=0.6$(a) and ${{M}_{t0}}=0.9$(b).}
\end{figure}

Since compressible turbulent flows own significant features of the fluctuations in thermodynamic variables, it is of interest to study their spectra under the presence of thermal fluctuations. Figure \ref{fig8} shows the spectra of $T$, $n$ and $P$ obtained by USP and DNS simulations at $t=1.4{{T}_{e0}}$ for different cases, where the USP spectra grow quadratically with $k$ in the high wavenumber range, indicating the effect of thermal fluctuations. Following the previous discussions, we calculate the thermal fluctuation spectra with $F$ = 1 to obtain the crossover wavenumbers ${{k}_{T}}$, ${{k}_{n}}$ and ${{k}_{P}}$ for the real gas flows. It is interesting to observe that as ${{M}_{t}}$ increases, the ranges of variation for ${{k}_{T}}\eta$, ${{k}_{n}}\eta$ and ${{k}_{P}}\eta$ are (2.1, 3.1), (2.1, 3.3) and (2.1, 3.5), respectively, which are close to the aforementioned trend observed for $k_{c}^{c}\eta$. To further verify the coupling relationship between the spectra of thermodynamic variables and the spectrum of ${{\vec{u}}_{c}}$, in figure \ref{fig9} we present the corresponding USP results for the cases with ${{M}_{t0}}=0.6$ and ${{M}_{t0}}=0.9$, following the normalization rules that the integral of the spectrum over the entire wavenumber range is equal to 1. It can be seen that the spectra of all the thermodynamic variables show good agreement with ${{E}_{c}}(k)$, indicating that the spatial correlations of thermodynamic fluctuations are dominated by the compressible mode of the velocity field. It is worth noting that this phenomenon was also reported in our previous work where we simulated the 2D decaying isotropic turbulence using the DSMC method \citep{Ma2023thermal}.

\section{Thermal fluctuations and the turbulence predictability}
\label{sec:predict}
Our research above indicates that thermal fluctuations have a significant impact on the turbulent spectra at length scales comparable to the turbulent dissipation length scale. On the other hand, this suggests that the large-scale turbulence statistics are unaffected by thermal fluctuations. However, if we shift our focus to the turbulent flow field structures, the situation may be quite different. Considering the current experimental challenges in directly measuring thermal fluctuations in turbulent flows \citep{Bandak2022thermalPRE}, the initial state of thermal fluctuations remains unknown when we attempt to predict the subsequent evolution of turbulence. Due to the chaotic nature of turbulent flows, even tiny deviations in the initial thermal fluctuations may lead to the unpredictability of large-scale turbulence structures after a certain period of time \citep{Ruelle1979micro}.

In this section, we employ the USP method to study the predictability of compressible decaying isotropic turbulence for both 2D and 3D cases. The initial temperature and pressure of the simulation cases are consistent with those in the previous sections, and the other simulated parameters are shown in table \ref{predict_parameter}. Note that the initial Taylor Reynolds number has different definitions for 2D and 3D turbulent flows, as shown in (\ref{Mach_Re_2D}) and (\ref{Re_lamda3D}), respectively. Therefore, in table \ref{predict_parameter} we provide the initial global Reynolds number ${{{Re}}_{0}}$, which is given as
\begin{equation}
  R{{e}_{0}}=\frac{{{U}_{0}}L}{2\pi {{\nu }_{0}}},
\end{equation}
where $L$ is the side length of the simulation domain. Based on $L$, one can also calculate the initial global Knudsen numbers as $K{{n}_{0}}={{{\lambda }_{mic0}}}/{L}\;$.
\begin{table}
  \centering
    \begin{tabular}{*{10}{c}}
    %\hline
    Case & $K{{n}_{0}}$  & ${{Re}_{0}}$  & ${{M}_{t0}}$ & Resolution & $\left\langle {{N}_{p}} \right\rangle$ & $F$ & ${\Delta t}/{{{\tau }_{mic0}}}\;$ &  ${\Delta {{L}_{cell}}}/{{{\lambda }_{mic0}}}\;$ & ${{k}_{\max }}{{\eta }_{\varOmega 0}}$ (${{k}_{\max }}{{\eta }_{0}}$) \\
    \midrule
    2D  &0.000125 &1590 & 1.0  &  ${{2048}^{2}}$	&400	&6194	 &2	 &3.9	 &18.7  	     \\
    3D  &0.00025  &954  & 1.2  &  ${{1024}^{3}}$	&25	  &7354	 &2	 &3.9	 &7.84  	     \\\hline
    \end{tabular}
    \caption{USP simulation parameters for the study of turbulence predictability.}
  \label{predict_parameter}
\end{table}

To investigate the effect of thermal fluctuations on the predictability of turbulence, note that the velocity field in a USP simulation is initialized as $\vec{u}_{0}^{\text{USP}}=\vec{u}_{0}^{\text{NS}}+\vec{u}_{0}^{\text{th}}$. Therefore, for both 2D and 3D cases, we can create an ensemble of realizations starting with the same $\vec{u}_{0}^{\text{NS}}$, but with different $\vec{u}_{0}^{\text{th}}$ generated using independent random number streams. This approach ensures that each realization initially differs from others solely due to small-scale thermal fluctuations.

\begin{figure}
  \centering
  \includegraphics[scale=0.45]{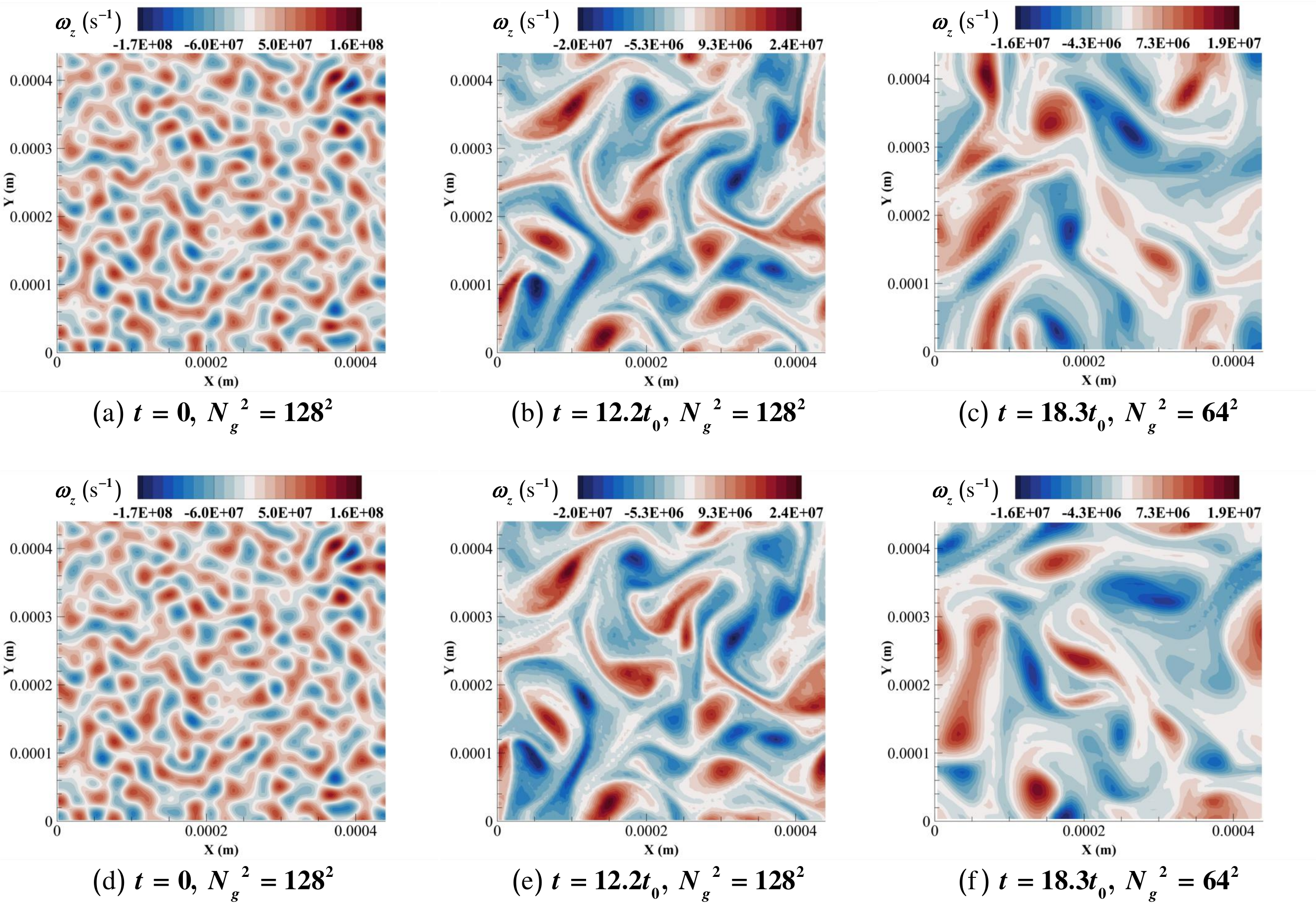}
  \caption{Temporal evolution of the vorticity field for 2D decaying turbulence with $R{{e}_{0}}$=1590 and ${{M}_{t0}}$=1. Panels (a-c) and (d-f) correspond to two realizations. The contours are plotted based on coarse-grained cells with a lower resolution (${{N}_{g}}^{2}$).}
\label{fig10}
\end{figure}

\begin{figure}
  \centering
  \includegraphics[scale=0.45]{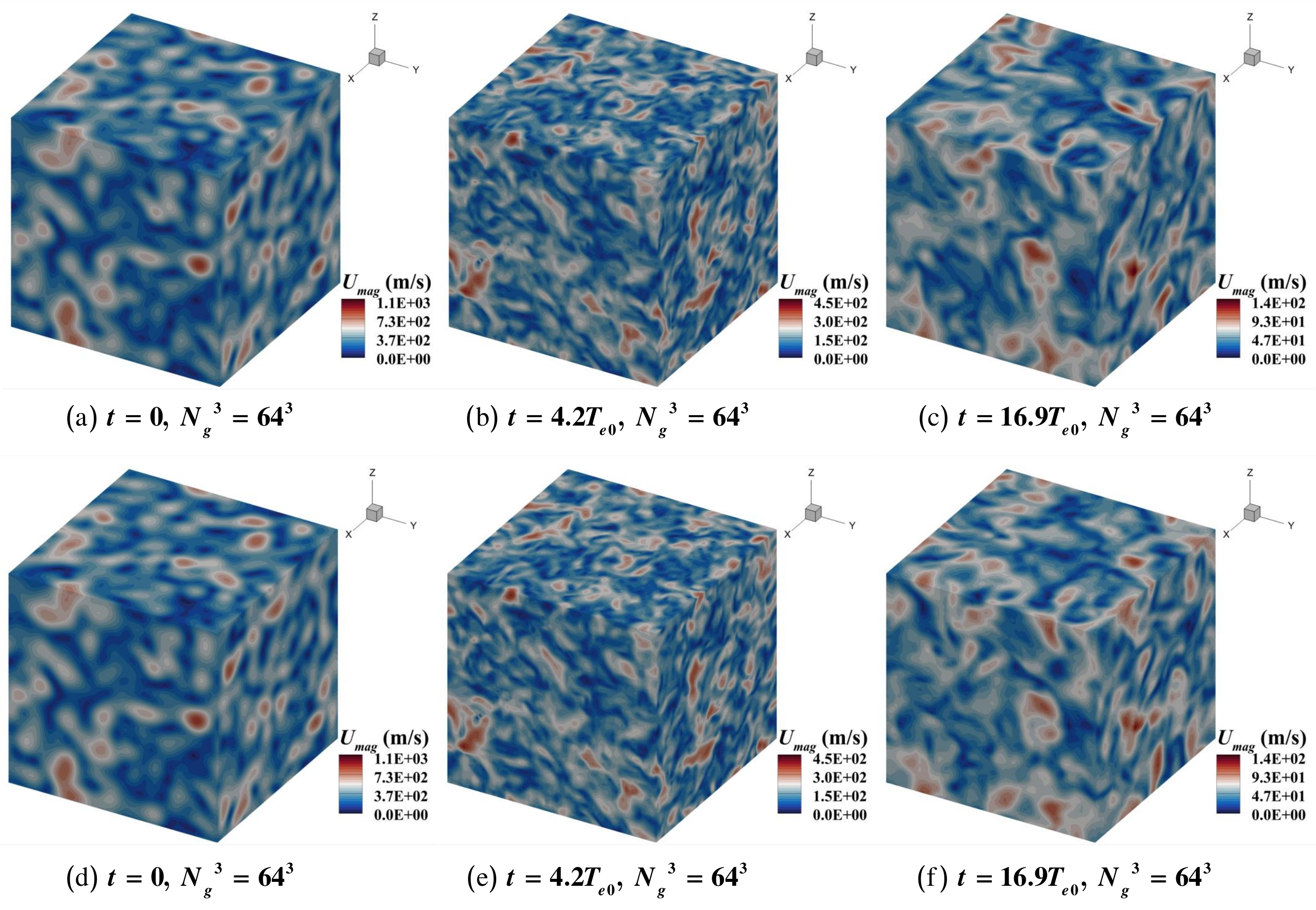}
  \caption{Temporal evolution of the velocity field for 3D decaying turbulence with $R{{e}_{0}}$=954 and ${{M}_{t0}}$=1.2. Panels (a-c) and (d-f) correspond to two realizations.The contours are plotted based on coarse-grained cells with a lower resolution (${{N}_{g}}^{3}$).}
\label{fig11}
\end{figure}

Figure \ref{fig10} shows the temporal evolution of the vorticity fields for two realizations of 2D decaying turbulence. Since we focus on the predictability of large-scale structures in turbulent flows, the contours are plotted based on “coarse-grained” cells with a lower resolution, i.e., the length of coarse-grained cells ${L}/{{{N}_{g}}}$ is much larger than the original cell length $\Delta {{L}_{cell}}$. As seen in figures \ref{fig10}(a) and (d), the vorticity fields of two realizations are identically the same at $t=0$, indicating that the thermal fluctuations have no effect on the initial large-scale turbulent structures. At $t=12.2{{t}_{0}}$, the vorticity fields of two realizations show slight differences (see figures \ref{fig10}(b) and (e)), and finally the vorticity fields show significant divergence at $t=18.3{{t}_{0}}$ (see figures \ref{fig10}(c) and (f)). For the 3D case, a similar phenomenon is observed in figure \ref{fig11}, where we compare the velocity magnitudes ${{U}_{mag}}$ of two realizations. At $t=16.9{{T}_{e0}}$, which corresponds to the ending time of 3D simulations, the velocity fields of the two realizations show observable differences (see figures \ref{fig11}(c) and (f)). Therefore, it can be concluded that the initial differences in small-scale thermal fluctuations will lead to the unpredictability of large-scale structures of turbulence after a certain period of time.

\begin{figure}
  \centering
  \subfloat{
      \includegraphics[scale=0.5]{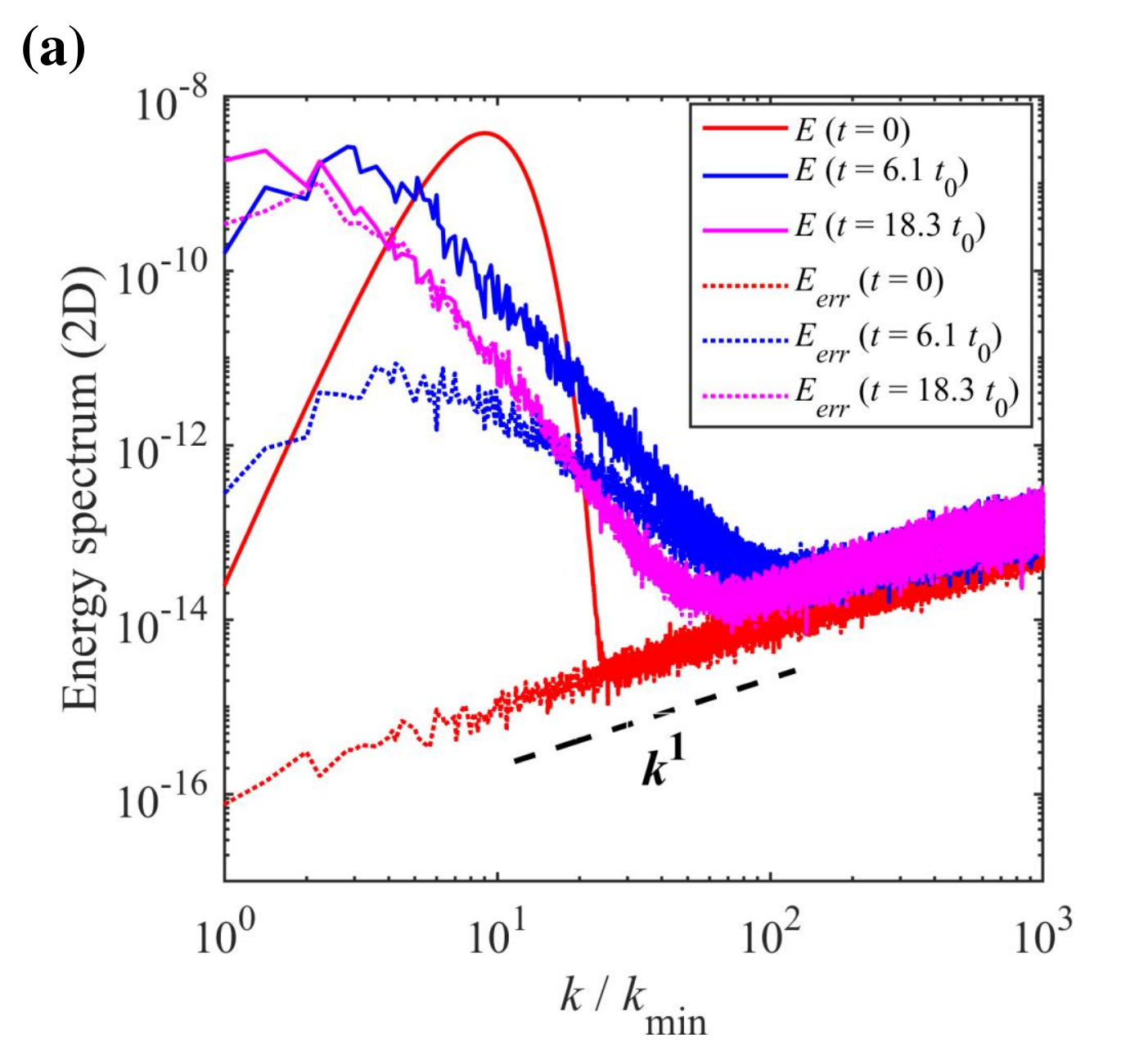}}
  \subfloat{
      \includegraphics[scale=0.5]{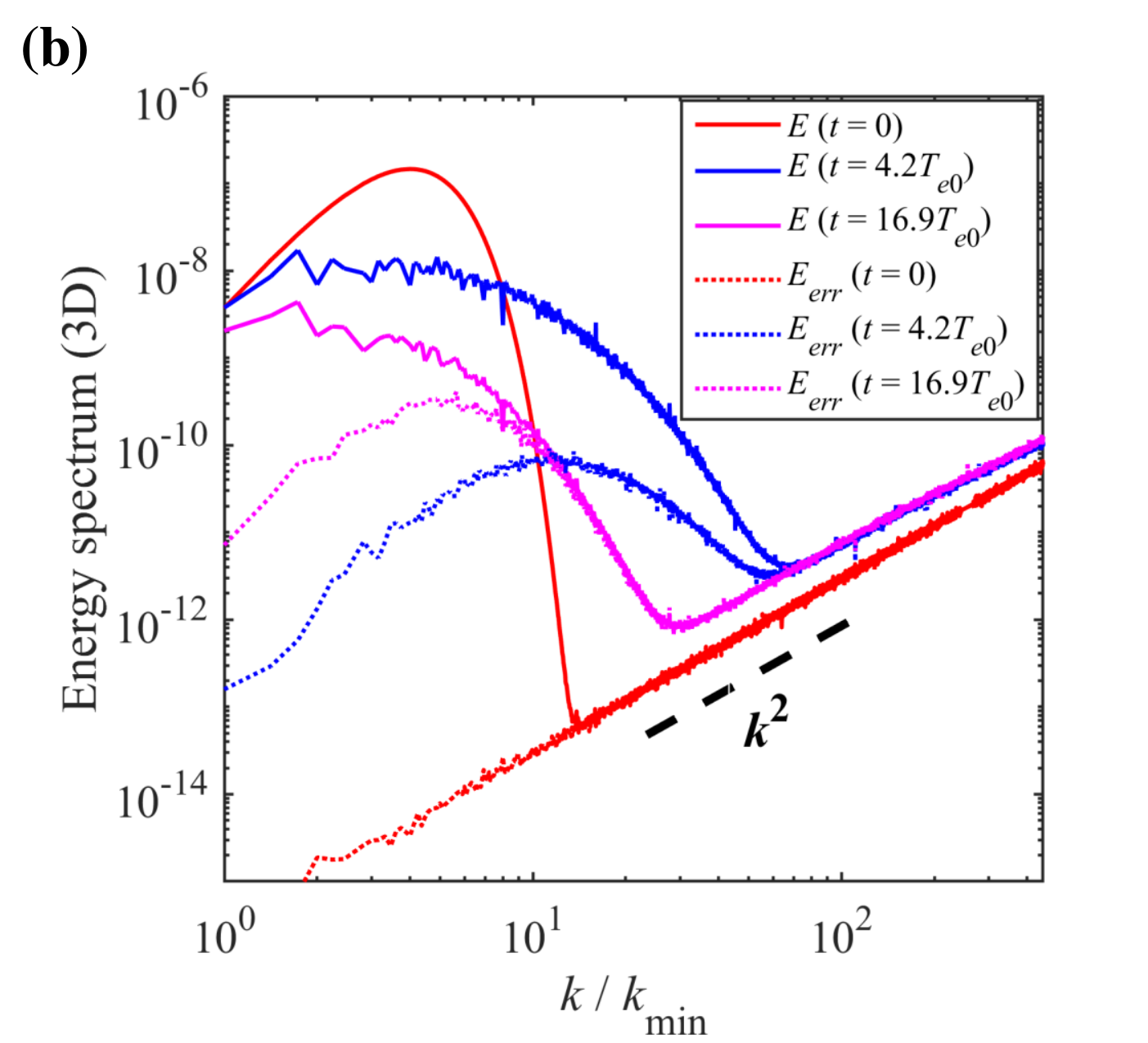}}
  \caption{\label{fig12}Energy spectra of the original velocity field and the error velocity field at different time points, for 2D (a) and 3D (b) cases.}
\end{figure}

To quantify the divergence between different flow realizations, we define the local error velocity field as \citep{Boffetta1997predict2D,Boffetta2017predict}
\begin{equation}
  {{\vec{u}}_{error}}(\vec{r},t)=\frac{1}{\sqrt{2}}({{\vec{u}}_{1}}(\vec{r},t)-{{\vec{u}}_{2}}(\vec{r},t)),
\end{equation}
where ${{\vec{u}}_{1}}$ and ${{\vec{u}}_{2}}$ represent the velocity fields of two independent realizations. In figure \ref{fig12}, we compare the energy spectra of original velocity fields and error velocity fields at different time points for both 2D and 3D cases. The results are obtained by taking the ensemble average of three independent realizations. Since $\vec{u}_{0}^{\text{th}}$ in each pair of realizations can be considered as two sets of independent Gaussian random variables \citep{Landau1980equilibrium}, the initial error velocity field can be regarded as a new thermal fluctuation field with the same Gaussian statistics. As shown in figure \ref{fig12}, the error spectra ${{E}_{err}}$ at $t$ = 0 grow linearly/quadratically with $k$ for 2D/3D cases, reflecting the basic features of thermal fluctuations. As time progresses, ${{E}_{err}}$ is still dominated by thermal fluctuations in the high wavenumber region, while gradually approaching $E$ in the low wavenumber region. This indicates that the initial errors of thermal fluctuations propagate to the larger scales. It is worth noting that, the “inverse cascade” of the error velocity field has also been observed in the previous studies based on the deterministic NS equations \citep{Olivier1986predict,Boffetta1997predict2D,Boffetta2001predict,Boffetta2017predict,Berera2018chaotic}, where the divergence of velocity field trajectories is achieved by initially introducing an artificial perturbation. Compared to these works, in our current research, the initial errors originate from thermal fluctuations, which are inherent properties of fluids. Furthermore, the influence of thermal fluctuations persists throughout the turbulent evolution process, rather than being limited to the initial moment.

\begin{figure}
  \centering
  \subfloat{
      \includegraphics[scale=0.5]{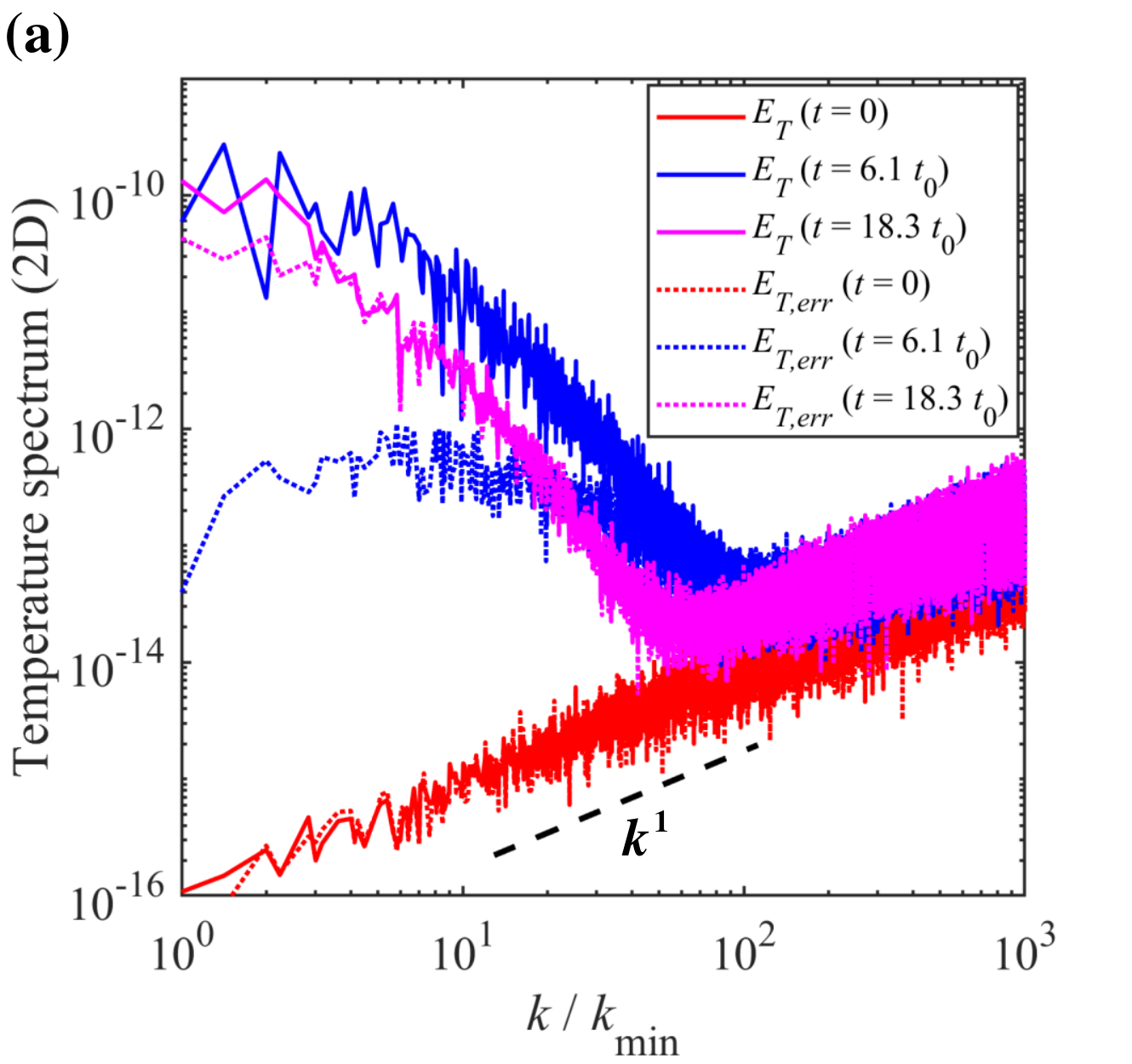}}
  \subfloat{
      \includegraphics[scale=0.5]{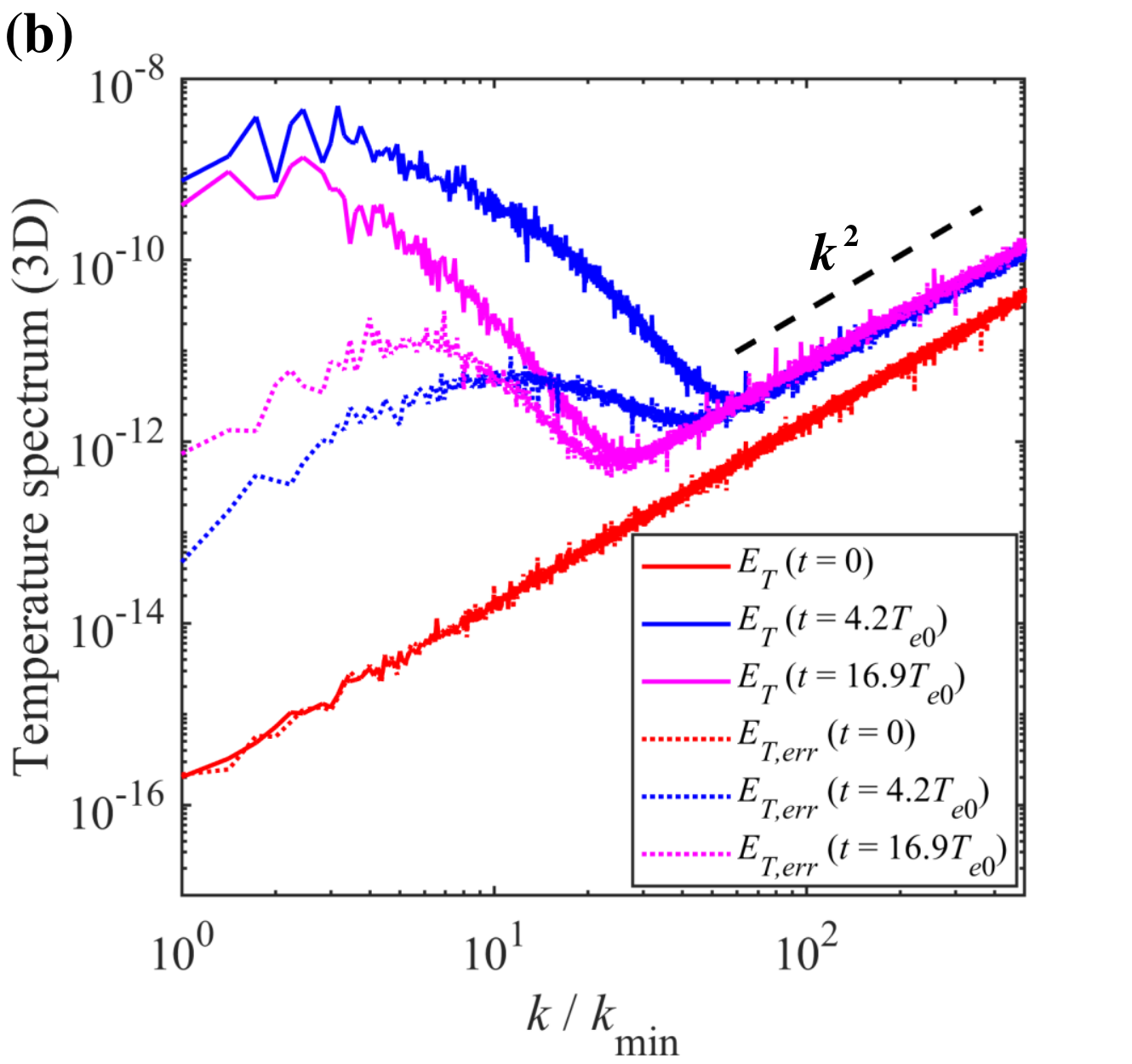}}
  \caption{\label{fig13}Spectra of the original temperature field and the error temperature field at different time points, for 2D (a) and 3D (b) cases.}
\end{figure}

\begin{figure}
  \centering
      \includegraphics[scale=0.6]{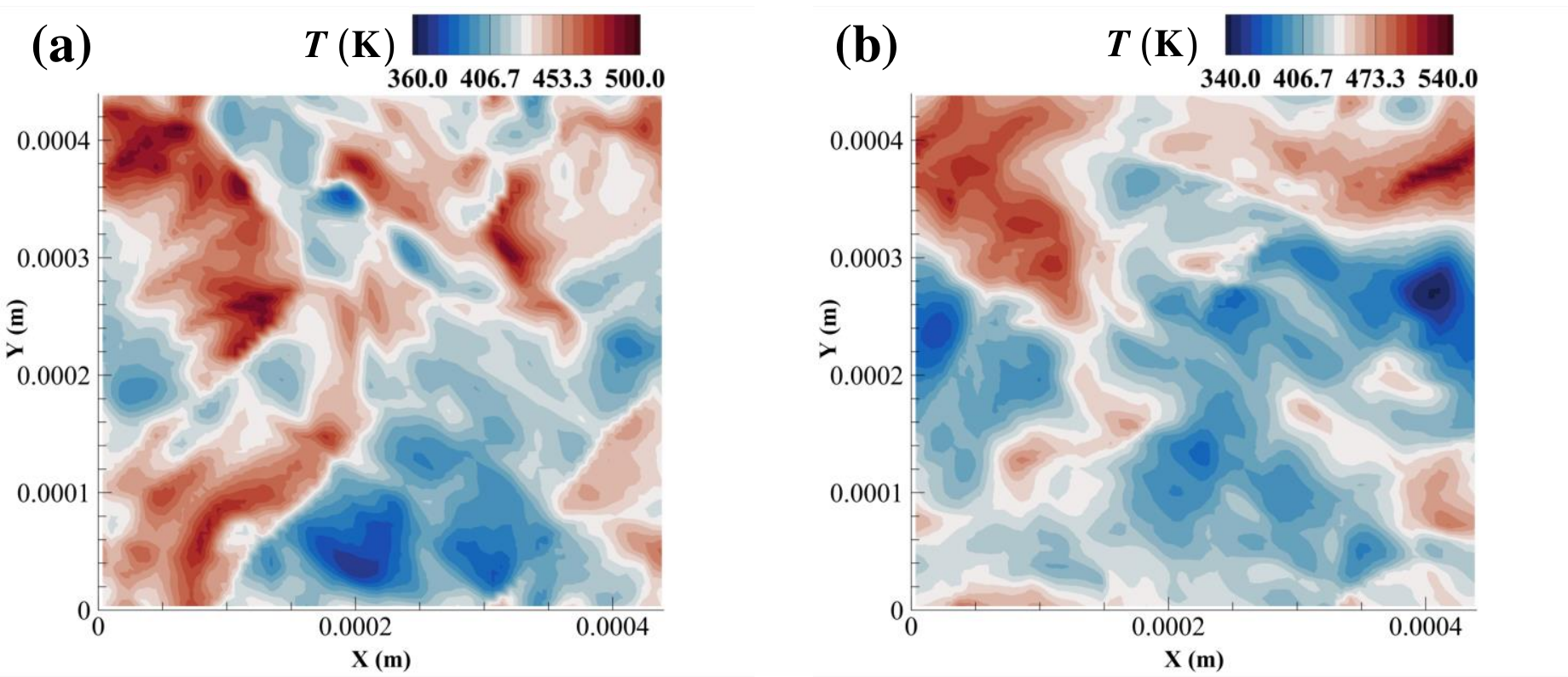}
  \caption{\label{fig14}Temperature field for 2D decaying turbulence with $R{{e}_{0}}$=1590 and ${{M}_{t0}}$=1, at $t$ = 18.3$t_0$. Panels (a) and (b) correspond to two realizations.}
\end{figure}

\begin{figure}
  \centering
      \includegraphics[scale=0.6]{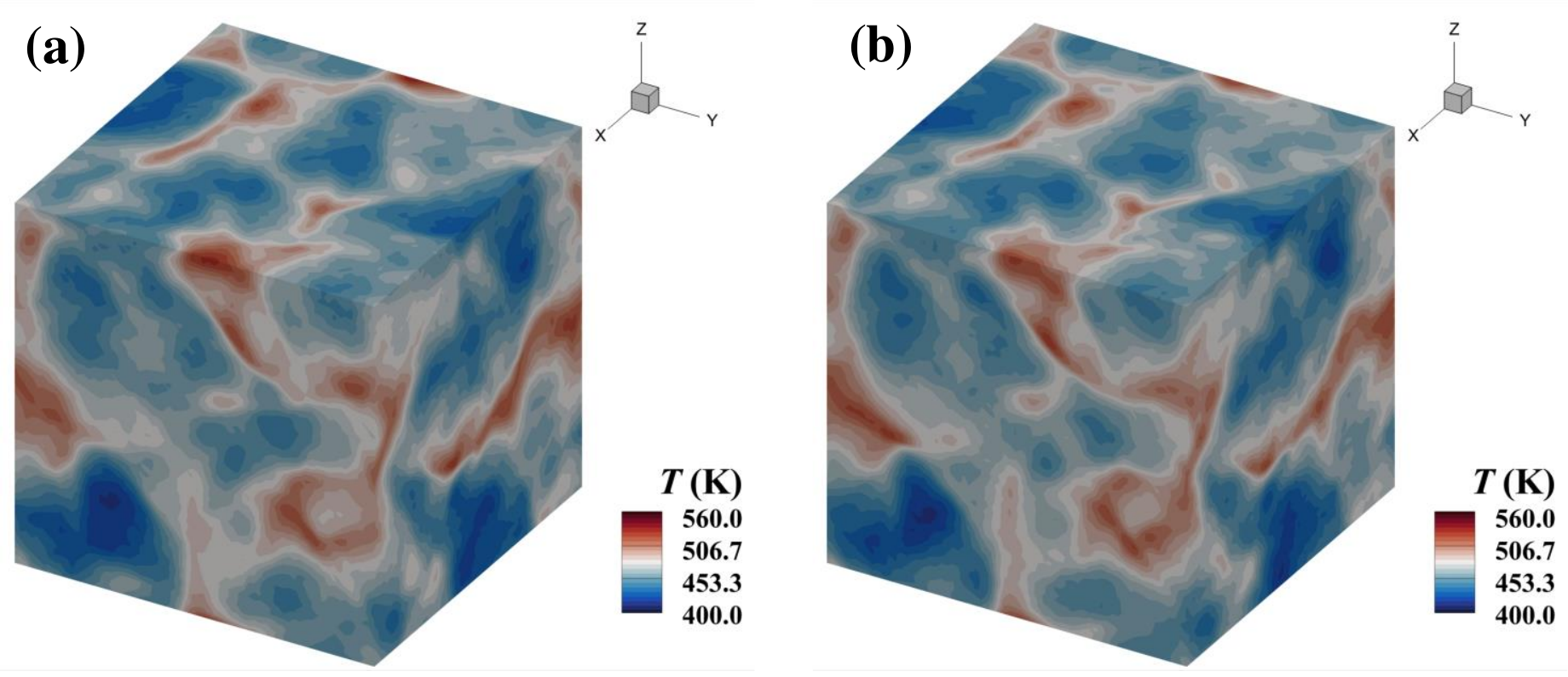}
  \caption{\label{fig15}Temperature field for 3D decaying turbulence with $R{{e}_{0}}$=954 and ${{M}_{t0}}$=1.2, at $t$ = 16.9${{T}_{e0}}$. Panels (a) and (b) correspond to two realizations.}
\end{figure}

In addition to the turbulent velocity field, we further investigate the effect of thermal fluctuations on the predictability of the turbulent thermodynamic field, and this aspect has not been reported in the literature. We define the local error fields of the thermodynamic variables as
\begin{equation}
  {{g}_{error}}(\vec{r},t)=\frac{1}{\sqrt{2}}({{g}_{1}}(\vec{r},t)-{{g}_{2}}(\vec{r},t)),
\end{equation}
where $g$ stands for $T$, $n$ and $P$. Figure \ref{fig13} presents a comparison between the spectra of the original temperature fields and the error temperature fields for both 2D and 3D cases. Similar comparisons can also be made for number density and pressure. At the beginning of USP simulations, the fluctuations of thermodynamic variables are solely attributed to thermal fluctuations, resulting in a complete coincidence between ${{E}_{T}}$ and ${{E}_{T,err}}$. Then, the compressibility of turbulence causes a rapid amplification of ${{E}_{T}}$ to a high magnitude in the low wavenumber region, while ${{E}_{T,err}}$ requires a considerably longer time to approach ${{E}_{T}}$. In figures \ref{fig14} and \ref{fig15}, we present the temperature fields of two realizations at the end of the USP simulations for 2D and 3D cases, respectively. Compared to the 3D cases, the differences between the two realizations are more pronounced for the 2D cases, which can be supported by figure \ref{fig13}(a) where ${{E}_{T,err}}$ almost coincides with ${{E}_{T}}$ at the end of the simulation.

\begin{figure}
  \centering
  \subfloat{
      \includegraphics[scale=0.5]{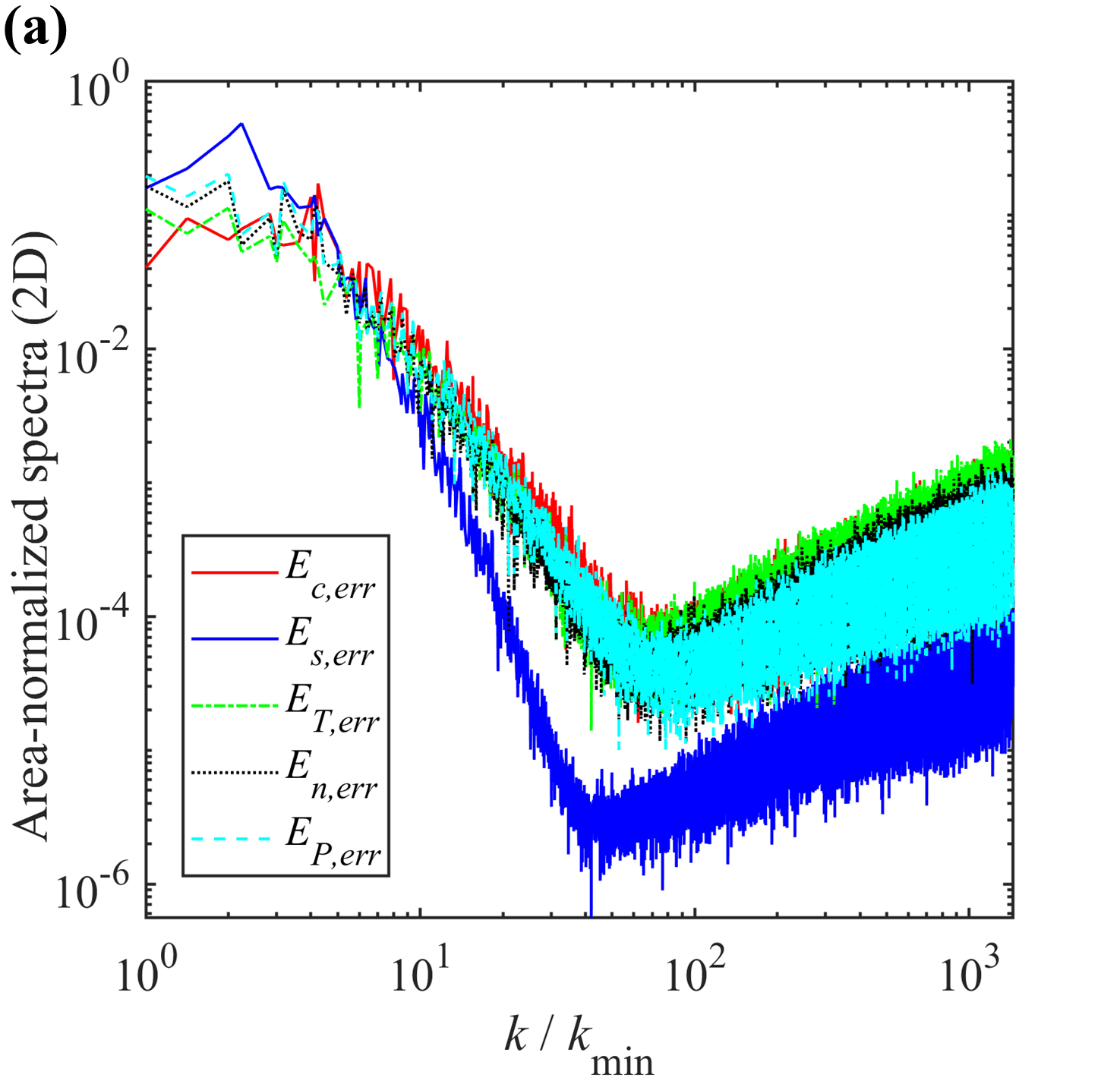}}
  \subfloat{
      \includegraphics[scale=0.5]{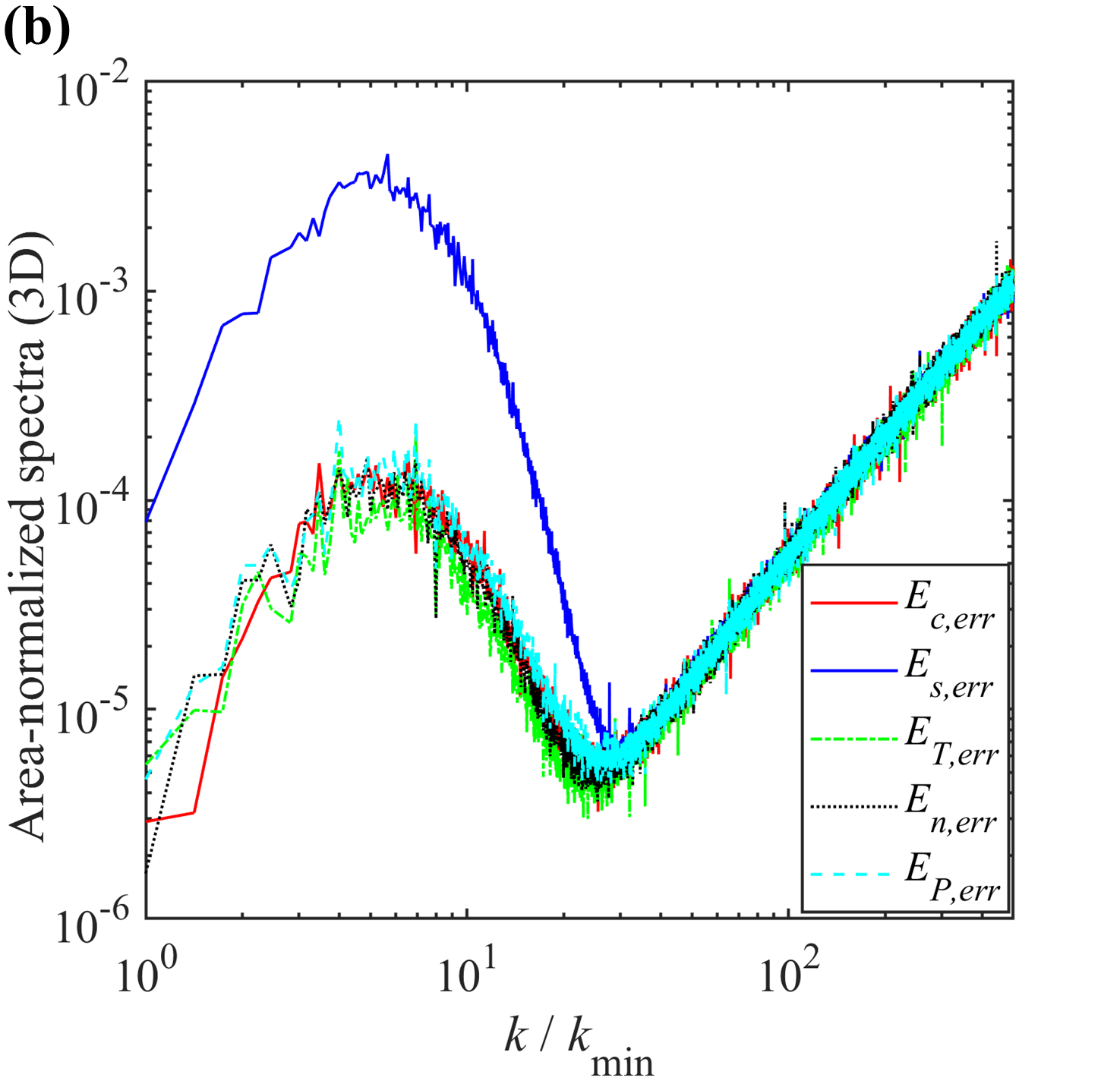}}
  \caption{\label{fig16}(a) Normalized error spectra of compressible velocity component, solenoidal velocity component, temperature, number density and pressure for 2D decaying turbulence at $t$ = 18.3$t_0$. (b) The error spectra corresponding to the 3D decaying turbulence at t = 16.9${{T}_{e0}}$.}
\end{figure}

In §\,\ref{sec:3D}, we have discussed the coupling relationship between turbulent thermodynamic variables and the compressible velocity component (see figure \ref{fig9}). To see whether this relationship holds for the turbulent error field, we compare the corresponding area-normalized spectra for both 2D and 3D cases in figure \ref{fig16}. It is evident that the error spectra of all the thermodynamic variables are in good agreement with the spectrum of the compressible component of the error velocity field. This suggests that in decaying compressible turbulence, the predictabilities of the thermodynamic variables are closely related to the predictability of the compressible velocity component.
\section{Concluding remarks}
\label{sec:conclusions}
In this work, we employed the USP method to numerically investigate the effects of thermal fluctuations on compressible decaying isotropic turbulence. Compared to the traditional molecular methods such as DSMC, USP can be applied with much larger time steps and cell sizes as it couples the effects of molecular movements and collisions. 

In both 2D and 3D simulations, it is found that the turbulent spectra of velocity and thermodynamic variables are greatly affected by thermal fluctuations in the high wavenumber range. The wavenumber scaling law of the thermal fluctuation spectra depends on the spatial dimension $d$ as ${{k}^{(d-1)}}$. By applying the Helmholtz decomposition to the velocity field, we show that the thermal fluctuation spectra of solenoidal and compressible velocity components (i.e., ${{E}_{s}}(k)$ and ${{E}_{c}}(k)$) follow an energy ratio of 1:1 for 2D cases, while this ratio changes to 2:1 for 3D cases.

For 3D decaying turbulence, a comparative study was conducted between the USP method and the DNS method based on the deterministic NS equations. The crossover wavenumbers were determined as the intersections between DNS spectra and thermal fluctuation spectra. The results show that thermal fluctuations dominate the turbulent spectra below length scales comparable to the Kolmogorov length scale $\eta$, which shows good agreement with the previous studies \citep{McMullen2022NSturbulence,Bell2022thermal}. Furthermore, it is observed that the crossover wavenumbers of thermodynamic spectra increase with ${M}_{t}$ following the same trend as the crossover wavenumber of ${{E}_{c}}(k)$, indicating the coupling relationship between thermodynamic fluctuations and the compressible mode of the velocity field. 

In addition to the turbulent spectra, our results demonstrate that thermal fluctuations also play an important role in the predictability of turbulence. Specifically, with initial differences attributed solely to small-scale thermal fluctuations, different flow realizations exhibit large-scale divergences in velocity and thermodynamic fields after a certain period of time. By calculating the error spectra between flow realizations, our study reveals the "inverse error cascades" for velocity and thermodynamic variables. Moreover, our results suggest a strong correlation between the predictabilities of thermodynamic variables and the predictability of the compressible velocity component.

In this study, we focused on the effects of thermal fluctuations on homogeneous isotropic turbulence, but we expect thermal fluctuations to be important for other turbulent flow scenarios, such as turbulent mixing \citep{Eyink2022mixing} and laminar–turbulent transition \citep{Lin2016intrinsicRandom,Luchini2016Wing}. For the predictability of turbulence, further research is required to investigate the error growth rates associated with different Reynolds numbers. For such scenarios and others, the USP method provides a powerful numerical tool for future work.

\backsection[Acknowledgements]{The authors thank Prof. Fei Fei, Dr. Chengxi Zhao and Yuandong Chen for helpful discussions about this work.}

\backsection[Funding]{This work was supported by the National Natural Science Foundation of China (Grant Nos. 92052104 and 12272028). Part of the results were obtained on the Zhejiang Super Cloud Computing Center Zone M6, and others were obtained on the Beijing Super Cloud Computing Center Zone A.}

\backsection[Declaration of interests]{The authors report no conflict of interest.}

\bibliographystyle{jfm}
\bibliography{reference_qihan}

\end{document}